\documentclass[journal=jacsat,manuscript=article]{achemso}
\usepackage[utf8]{inputenc}
\usepackage[T1]{fontenc}
\usepackage[version=3]{mhchem} %
\usepackage[usenames,dvipsnames]{xcolor}
\usepackage{lipsum}
\usepackage[normalem]{ulem}

\newcounter{Paragraph}
\usepackage{amssymb}
\usepackage{amsmath}
\usepackage{graphicx}
\usepackage{dcolumn}
\usepackage{xcolor}
\usepackage{fancyhdr}
\usepackage[colorlinks,allcolors=black,citecolor=blue,urlcolor=blue]{hyperref}

\newcommand{\newinfo}[1]{{\color{black}{#1}}}

\title{Breaking the Air–Water Paradigm: Ion Behavior at Hydrophobic Solid–Water Interfaces}

\author{Xavier R. Advincula}
\affiliation{Yusuf Hamied Department of Chemistry, University of Cambridge, Lensfield Road, Cambridge, CB2 1EW, UK}
\alsoaffiliation{Cavendish Laboratory, Department of Physics, University of Cambridge, Cambridge, CB3 0HE, UK}
\alsoaffiliation{Lennard-Jones Centre, University of Cambridge, Trinity Ln, Cambridge, CB2 1TN, UK}

\author{Kara D. Fong}
\affiliation{Division of Chemistry and Chemical Engineering, California Institute of Technology, Pasadena CA 91125, USA}
\alsoaffiliation{Marcus Center for Theoretical Chemistry, California Institute of Technology, Pasadena CA 91125, USA}

\author{Yongkang Wang}
\affiliation{Max Planck Institute for Polymer Research, Ackermannweg 10, 55128 Mainz, Germany}

\author{Christoph Schran}
\affiliation{Cavendish Laboratory, Department of Physics, University of Cambridge, Cambridge, CB3 0HE, UK}
\alsoaffiliation{Lennard-Jones Centre, University of Cambridge, Trinity Ln, Cambridge, CB2 1TN, UK}

\author{Mischa Bonn}
\affiliation{Max Planck Institute for Polymer Research, Ackermannweg 10, 55128 Mainz, Germany}

\author{Angelos Michaelides}
\email{am452@cam.ac.uk}
\affiliation{Yusuf Hamied Department of Chemistry, University of Cambridge, Lensfield Road, Cambridge, CB2 1EW, UK}
\alsoaffiliation{Lennard-Jones Centre, University of Cambridge, Trinity Ln, Cambridge, CB2 1TN, UK}

\author{Yair Litman}
\affiliation{Max Planck Institute for Polymer Research, Ackermannweg 10, 55128 Mainz, Germany}
\email{litmany@mpip-mainz.mpg.de}

\begin{document}
\begin{tocentry}
\vspace{2mm}
\includegraphics[width=\textwidth]{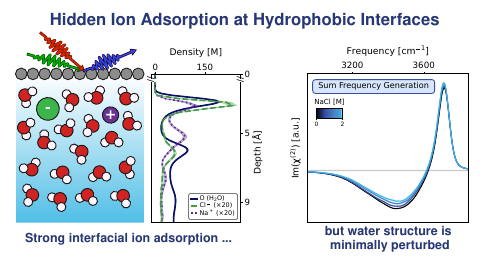}
\end{tocentry}

\begin{abstract}

Hydrophobic solid–water interfaces underpin processes in nanofluidics, electrochemistry, and energy technologies. 
Microscopic insights into these systems are often inferred from our understanding of the air–water interface, which is assumed to exhibit similar behavior.
Here, we challenge this paradigm by combining heterodyne-detected vibrational sum-frequency generation spectroscopy with machine-learning molecular dynamics simulations at first-principles accuracy to investigate the graphene–NaCl(aq) interface as a prototypical hydrophobic solid–water system.
Spectroscopic results suggest that ions have a minimal effect on the structure of the interfacial water, while simulations reveal that Na$^{+}$ and Cl$^{-}$ accumulate densely at the surface. 
Together, these findings reveal a new adsorption mechanism that departs from the established air–water interface paradigm, where interfacial ion adsorption is typically associated with, and often detected through, pronounced alteration of the interfacial water alignment and orientation. 
This difference arises because ions cannot penetrate the solid boundary and reside at a similar depth as the interfacial water molecules.
As a consequence, large ion populations can be accommodated within the extended two-dimensional hydrogen-bond network at the interface, causing only minor local distortions but significant changes to its longer-range connectivity.
These results reveal a distinct mechanism of electrolyte organization at aqueous–carbon interfaces, relevant to energy applications, where performance is highly sensitive to the local organization of interfacial water.

\end{abstract}

\maketitle

\section*{Introduction}

Understanding how ions interact in water near hydrophobic interfaces has puzzled scientists for decades~\cite{Chandler2005,Tian_PNAS_2009,Vasudevan_PNAS_2014} and it lies at the heart of technologies ranging from water purification and desalination~\cite{desalination_1, desalination_2} and atmospheric chemistry~\cite{atmos_1, atmos_2, brooksi_2025} to biomimetic design~\cite{BHUSHAN20111} and electrochemical energy storage~\cite{clare_2016, forse_nanop_2024}.

\begin{figure*}
    \centering
    \includegraphics[width=\textwidth]{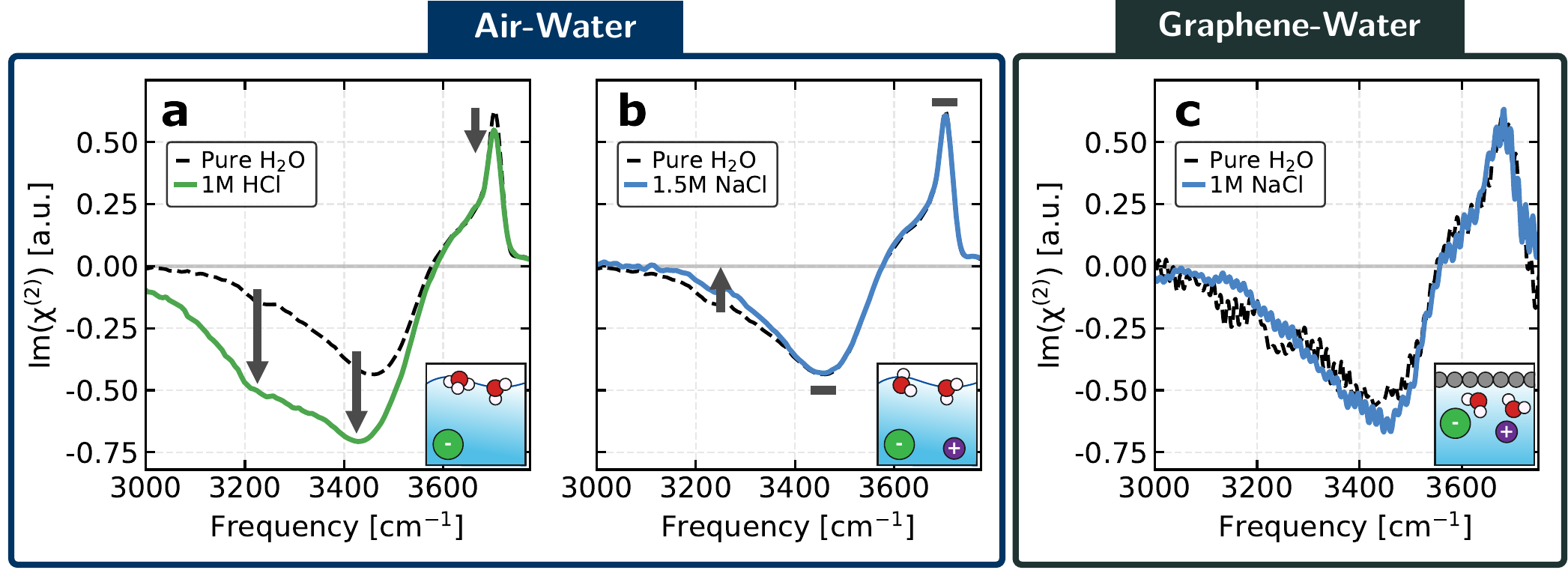}
    \caption{\textbf{Experimental HD-VSFG spectra of electrolyte solutions at air-liquid and solid-liquid interfaces.}
    HD-VSFG spectra at room temperature for (a) 1.0~M HCl at the air–water interface, (b) 1.5~M NaCl at the air–water interface, and (c) 1.0~M NaCl at the graphene–water interface.
    In each panel, the corresponding pure-water spectrum for the same interface is shown as a reference.
    (a) and (b) are adapted with permission from Litman \textit{et al.}~\cite{litman_surface_2024}. 
    The arrows indicate the direction of spectral changes with respect to the pure water spectrum, while gray rectangles mark regions where changes are negligible.
    The accompanying illustrations schematically indicate the presence or absence of ions at each interface.
    }
    \label{fig:fig1}
\end{figure*}

In this context, a significant body of literature has asserted that the water–air interface is the most common and simple aqueous interface and serves as a reference system to understand water at hydrophobic surfaces~\cite{Willard_JCP_2014,Bonn_ANIE_2015_review,Sedlmeier2008,Du_Science_1994,ohto_gra_hbn_2018}, and more specifically, that ion distributions at hydrophobic solid–water interfaces closely mirror those at the air-water boundary.
For example, Koelsch \textit{et al.}~\cite{koelsch_2007} reasoned that key properties of hydrophobic aqueous interfaces, including ion exclusion effects and hydration structure, are fundamentally similar to those observed at the air-water interface, attributing this to a generic hydrophobic effect that governs both environments.
Cui \textit{et al.}~\cite{cui_protein_hydro_2014} extended this analogy by demonstrating, through simulations, that the potential of mean force (PMF) profiles for chloride and iodide ions at both (hydrophobic) protein and air interfaces were nearly indistinguishable, thereby suggesting that molecular-level organization of ions and water is essentially shared between the two types of surfaces.
Furthermore, seminal works by the Saykally and Geissler groups~\cite{scox_pnas_2017,Devlin_pnas_2022} used nonlinear spectroscopic techniques and molecular simulations and determined that SCN$^-$ presents a very similar Gibbs adsorption free energy at the air-water, graphene-water, and toluene-water interfaces, even though different mechanisms might be at play.
More recently, Scalfi \textit{et al.}~\cite{scalfi_2024} reported that force-field and \textit{ab initio} simulations yield closely aligned PMF profiles for ions at both air and graphene interfaces, reinforcing the widespread notion that these interfaces are functionally analogous for ion adsorption phenomena.
Collectively, these studies suggest that the organization of interfacial water at the water–hydrophobic interface is characterized by relatively weak water–hydrophobe interactions and is dominated by strong water–water interactions.
These interactions maximize hydrogen bonding and promote the formation of a two-dimensional network, which underlies the anomalously high surface tension observed for water~\cite{Dannenberg_1997,pezzotti_2021_finger}.
In this canonical paradigm, the air-water interface can thus be regarded as the limiting case, where water–hydrophobe interactions vanish.
Accordingly, the presence of ions at hydrophobic interfaces is expected to disrupt the interfacial hydrogen-bond network as they do for the air-water interface~\cite{chiang_2020, das_ez_2020, sfg_proton_laage_2024, litman_surface_2024}.

In this work, we show that the canonical view of ion behavior at air-water interfaces does not necessarily apply to solid–water boundaries.
Ions can accumulate robustly at hydrophobic carbons without substantially perturbing the local structure of interfacial water.
To demonstrate this, we combine heterodyne-detected vibrational sum-frequency generation (HD-VSFG) with machine-learning molecular dynamics simulations.
\newinfo{While simple electrolytes such as NaCl exhibit pronounced stratification and ion-depleted surface layers at air–water interfaces~\cite{litman_surface_2024,netz_2012_rev, paesani_halide_air_wat_2025}, our results reveal that the water–graphene interface can host densely adsorbed ions while leaving the hydrogen-bond network largely intact.}
This marked contrast highlights that ion adsorption onto extended hydrophobic solids is primarily driven by subtle hydration structuring rather than by the direct disruption of interfacial water observed at air–water boundaries.
Thus, paradigms developed for the air-water interface cannot be simply transferred to solid–water interfaces.

\section*{Results}

\subsection*{Experimental VSFG Spectrum of NaCl(aq)/Graphene Is Indistinguishable from That of NaCl(aq)/Air}

To experimentally probe interfacial water structure in electrolyte solutions, VSFG spectroscopy has proven invaluable, as it is intrinsically surface-specific and highly sensitive to the orientation of interfacial water molecules~\cite{shen_og_1989_nature, shen_science_1994, bonn_science_2014, air_wat_sfg_paesani_2016, shen_2016_prl, saykally_sfg_2018, bonn_rev_2021, fellows_2024}.
Although atomic ions lack vibrational modes of their own, their presence can be inferred from the response of surrounding water molecules.
Phase-resolved VSFG, realized experimentally through HD-VSFG, extends this capability by providing direct access to the imaginary part of the nonlinear susceptibility, $\mathrm{Im}(\chi^{(2)})$, thereby unambiguously isolating the resonant vibrational response of interfacial water from the non-resonant responses~\cite{benderskii_2008, tahara_2013_rev, tahara_2015, wang_ange_2024}.
Moreover, the sign of the $\mathrm{Im}(\chi^{(2)})$ spectra reflects the net orientation of O--H bonds: a positive sign indicates O--H pointing toward the interface (away from the liquid), whereas a negative signal indicates downward orientation into the bulk~\cite{tahara_2008_up_down}.

\begin{figure*}
    \centering
    \includegraphics[width=\textwidth]{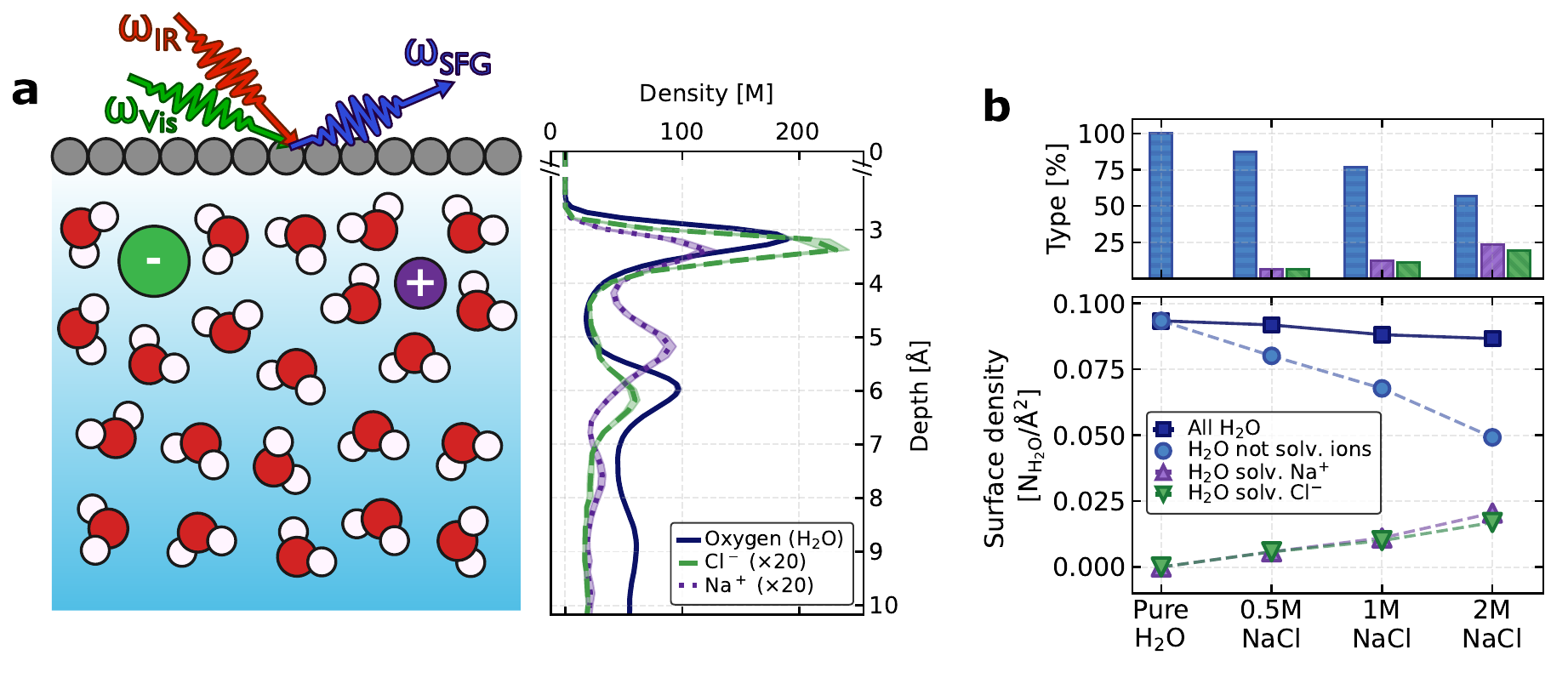}
    \caption{\textbf{Molecular structure of the graphene–NaCl(aq) interface at varying concentrations.}
    \newinfo{
    (a) Schematic illustration of the system studied, along with the density profiles of the water oxygen atoms, Cl$^{-}$ ions, and Na$^{+}$ ions at the graphene–NaCl(aq) interface for a 2~M NaCl solution.
    (b) Total number of interfacial water molecules, classified according to whether they solvate no ions, Na$^{+}$, or Cl$^{-}$.
    The accompanying bar plots show the percentage contribution of each type.
    Interfacial water molecules were defined as those located between the graphene surface and the first minimum of the water oxygen density profile (see Figure~S1), located at approximately 4.5~{\AA} from the surface.
    }
    }
    \label{fig:fig2}
\end{figure*}

We first consider the well-established case of HCl solutions, where surface-active protons adsorb strongly at the air–water interface~\cite{chiang_2020,das_ez_2020,sfg_proton_laage_2024}, disrupting the interfacial hydrogen-bond network.
As a result, the $\mathrm{Im}(\chi^{(2)})$ spectra change dramatically compared to pure water, with pronounced reshaping of the 3,000–3,600~cm$^{-1}$ region, arising from hydrogen–bonded O--H stretches, and suppression of the free O–H peak, corresponding to non-hydrogen-bonded O--H stretches. (Figure~\ref{fig:fig1}a)~\cite{das_ez_2020,chiang_2020}.
In contrast, when ions are excluded from the surface, as in NaCl solutions, the spectra remain largely similar to that of pure water (Figure~\ref{fig:fig1}b)~\cite{litman_surface_2024}.

In Figure~\ref{fig:fig1}c, we present the experimental HD-VSFG $\mathrm{Im}(\chi^{(2)})$ spectra for a prototypical hydrophobic solid–water interface, the focus of this work, the graphene-NaCl(aq) interface.
The spectrum displays a broad negative band centered between 3,200--3,550~cm$^{-1}$, corresponding to hydrogen-bonded O--H stretches, and a narrow positive peak above 3,600~cm$^{-1}$ associated with the dangling O–H of interfacial water, consistent with previous reports~\cite{wang_ange_2024,xu_structure_2023}, characteristic of a hydrophobic interface.

Compared to air–water measurements (Figures~\ref{fig:fig1}a,b), HD-VSFG at graphene–liquid interfaces is experimentally challenging and generally yields a lower signal-to-noise ratio, as seen in the noisier traces of Figure~\ref{fig:fig1}c.
Within the experimental uncertainty, both the hydrogen-bonded band and the dangling O--H peak exhibit indistinguishable changes upon NaCl addition at the graphene–water interface.
Such weak perturbations might appear to indicate only minor ionic effects, similar to the subtle spectral changes observed at the air–water interface~\cite{Shultz_2000,shen_2008,shen_2011,heather_2014}.
However, some of us have shown that ions can in fact accumulate at graphitic solid–liquid interfaces~\cite{kara_pairing_2024}.
Reconciling this apparently paradoxical behavior requires going beyond what the experimental spectra alone can reveal.

\subsection*{Simulations Reproduce the Experimental VSFG Spectrum and Reveal Interfacial Ion Enrichment}

Figure~\ref{fig:fig2}a illustrates the graphene-NaCl(aq) interface studied here.
We examine pure water and three electrolyte concentrations (0.5, 1, and 2 M NaCl).
To achieve multi-nanosecond statistics with near first-principles accuracy, we performed machine learning-based molecular dynamics simulations with a potential trained on revPBE–D3(0) reference data (see Methods).

\begin{figure*}
    \centering
    \includegraphics[width=\textwidth]{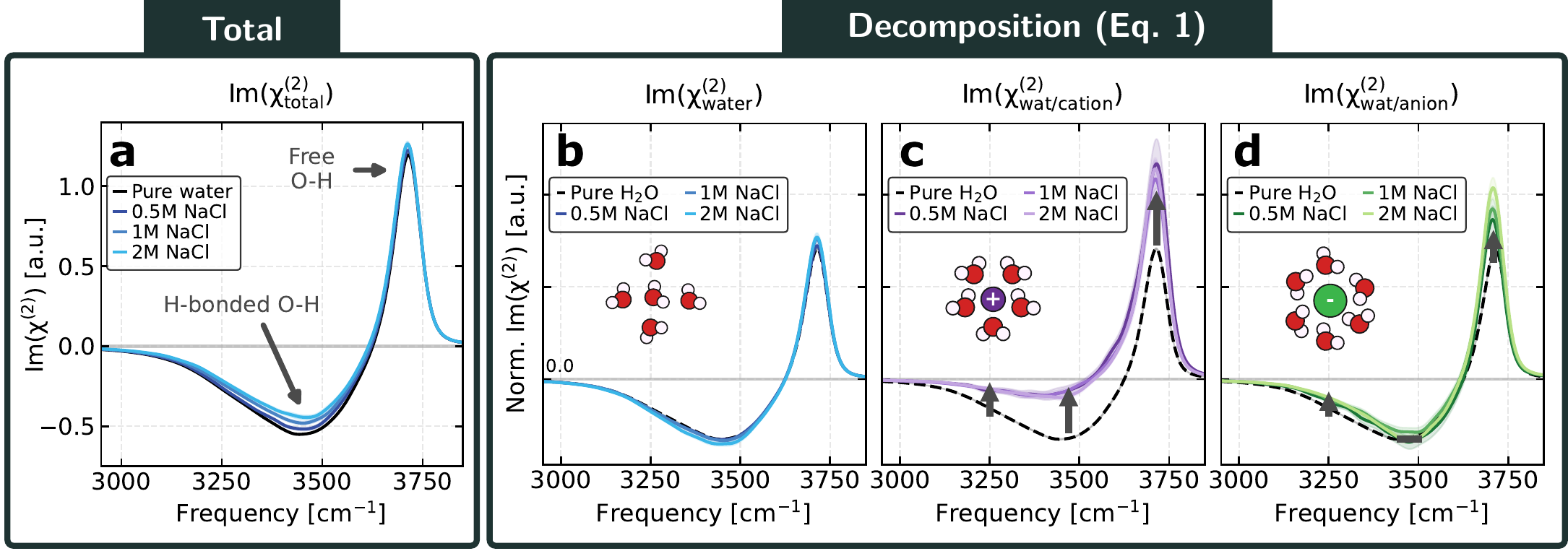}
    \caption{\textbf{Theoretical VSFG spectra of aqueous NaCl solutions at varying concentrations, with decomposition by O--H bond type.}
    \newinfo{
    (a) Theoretical $\mathrm{Im}(\chi^{(2)})$ spectra for the graphene–NaCl(aq) interface at NaCl concentrations of 0, 0.5, 1, and 2~M.
    (b--d) Decomposition of the spectra into contributions from 
    (b) water molecules not solvating ions,
    (c) Na$^{+}$-solvated waters, and 
    (d) Cl$^{-}$-solvated waters, obtained using the ssVVCF methodology~\cite{ssvcf_2015}. 
    The arrows indicate the direction of spectral changes with respect to pure water upon increasing salt concentration, while gray rectangles mark frequency regions where changes are negligible.
    In panels (b--d), the spectra are normalized by the number of interfacial water molecules of each type.}
    }
    \label{fig:fig3}
\end{figure*}

The associated Na$^+$ and Cl$^-$ ions are schematically depicted near the interface, reflecting the interfacial ion accumulation observed in the density profiles, consistent with our previous work~\cite{kara_pairing_2024}.
The oxygen density profiles exhibit markedly sharper interfacial layering than what is typically seen at the air–water interface~\cite{wc_2010, litman_surface_2024}, highlighting the stronger structural imprint of the solid substrate~\cite{angelos_tocci_2014, gra_wat_sfg_charged_paesani_2025}.
Figure~\ref{fig:fig2}b reports the surface density of water molecules in the topmost interfacial layer, classified according to their local environment: those not solvating ions, those solvating Na$^{+}$, and those solvating Cl$^{-}$.
With increasing salt concentration, the fraction of ion-solvating water rises (for Cl$^{-}$, from 0\% in pure water to 20\% at 2 M; for Na$^{+}$, from 0\% to 24\%).
At the same time, the population of non–ion-solvating water decreases (from 100\% in pure water to 56\% at 2~M), confirming ion adsorption at the interface and the reorganization of hydration shells.
Yet the total number of interfacial water molecules remains nearly constant, indicating that the hydrogen-bond network adapts to accommodate ion solvation without significantly disrupting the overall interfacial density and structure.

Despite the ion enrichment, the computed $\mathrm{Im}(\chi^{(2)})$ spectra reveal only subtle changes with increasing salt concentration (Figure~\ref{fig:fig3}a).
In the following, unless stated otherwise, we focus on the topmost interfacial water layer, as it provides the dominant contribution to the VSFG spectra, whereas the second layer affects the response only marginally (Figure~S4).
Upon addition of NaCl, we observe a moderate reduction in the magnitude of the hydrogen-bonded O--H band (less negative) and a slight enhancement of the dangling O--H peak (more positive).
As anticipated earlier, ions induce only minimal perturbation to the local structure of the interfacial water.
The graphene surface thus appears to buffer the structural impact of interfacial ions, leading to surprisingly subtle changes in the VSFG signal.

\subsection*{Cation Solvation Drives Subtle Changes in the VSFG Spectrum}

To understand the origin of the spectral changes, we decompose the total VSFG response into contributions from the local environments of the water molecules.
In this way, $\chi_{\textrm{total}}^{(2)}$ can be expressed as a linear combination of three distinct terms.
The first term, $\chi_{\textrm{water}}^{(2)}$, arises from water molecules not solvating ions.
The second term, $\chi_{\textrm{wat/cation}}^{(2)}$, comes from water molecules solvating Na$^{+}$.
The third arises, $\chi_{\textrm{wat/anion}}^{(2)}$, from water molecules solvating Cl$^{-}$~\cite{litman_surface_2024}:
\begin{equation}
    \chi_{\textrm{total}}^{(2)} = \chi_{\textrm{water}}^{(2)} + c \left(\chi_{\textrm{wat/cation}}^{(2)} + \chi_{\textrm{wat/anion}}^{(2)}\right), \label{eq:sfg_decomposition}
\end{equation}
where $c$ is a prefactor proportional to the NaCl concentration.
The spectra are then normalized by the number of interfacial water molecules in each class, so that we can focus on changes in spectral shape relative to the pure graphene-water interface.

\begin{figure*}
    \centering
    \includegraphics[width=\textwidth]{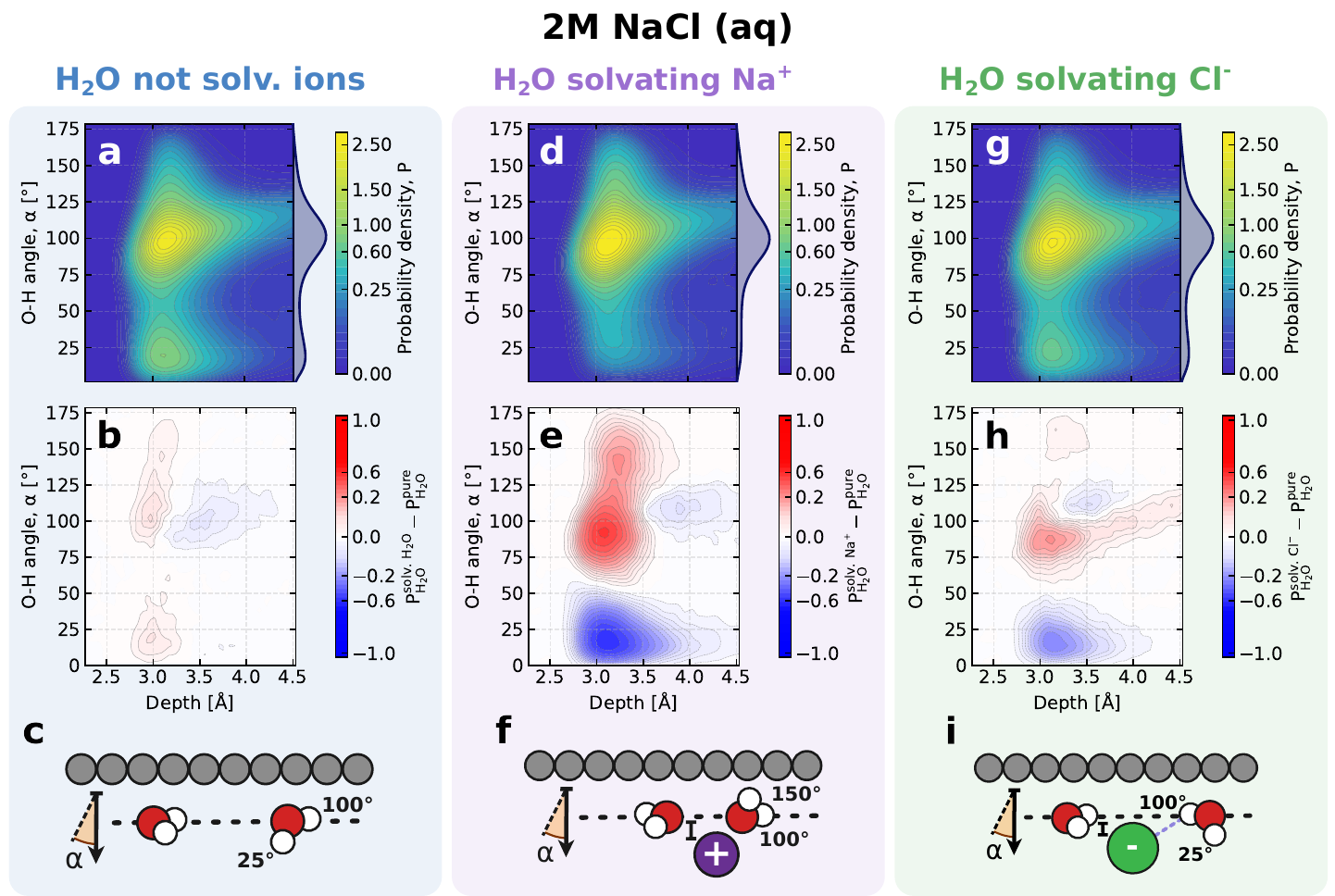}
    \caption{\textbf{Microscopic analyses of the graphene–NaCl(aq) interface, with molecular-level illustrations of how Na$^{+}$ and Cl$^{-}$ ions modify interfacial water structures for a 2~M NaCl solution.}
    (a) Probability distribution of O--H bond orientations in water molecules not solvating ions as a function of their depth, defined as the distance between the water oxygen and the graphene surface, and the angle relative to the surface normal.
    An angle of $0^\circ$ indicates that the O--H bond points toward the bulk solution.
    The accompanying profile shows the O--H angle distribution integrated over all depths, highlighting the most probable orientations.
    (b) Difference in the probability distribution shown in (a) relative to that of pure water at the graphene interface.
    Positive values indicate features that appear compared to pure water, while negative values indicate features that disappear compared to pure water.
    (c) Molecular-level depiction of water molecules solvating other water molecules at the graphene interface, including the definition of the O--H bond angle, $\alpha$.
    The horizontal dashed line indicates the position of the interfacial layer based on oxygen atom positions.
    The left water molecule illustrates the dominant in-plane orientation at the interface, whereas the right one represents the principal out-of-plane configuration.
    (d–f) Same as (a–c) but for water molecules solvating Na$^{+}$.
    (g–i) Same as (a–c) but for water molecules solvating Cl$^{-}$.
    For water molecules surrounding Cl$^{-}$, hydrogen bonding to the ion is indicated by the purple dotted line in (i).
    Note the relative depths of Na$^{+}$ and Cl$^{-}$ ions with respect to the horizontal dashed line.
    Results for additional concentrations are provided in Section~S2.
    }
    
    \label{fig:fig4}
\end{figure*}

In Figure~\ref{fig:fig3}b, we show the spectra from water molecules not solvating ions. 
We find that their spectral response remains essentially unchanged across all concentrations.
This supports our decomposition in Equation~\ref{eq:sfg_decomposition}, which can be viewed as a strong-solvation-shell model, where waters outside the ion solvation shells are assumed to remain unperturbed by the presence of salt.
We next examine the contribution from Na$^+$-solvating water molecules (Figure~\ref{fig:fig3}c).
Here, the differences are more pronounced.
In the hydrogen-bonded region, we observe a strong reduction in the signal magnitude (less negative).
In fact, the near-absence of signal suggests that these O--H bonds lie predominantly in-plane. 
Such orientations do not contribute to the response in the polarization combination used here, and therefore appear inactive in our spectra.
In the dangling O--H region, we find an enhanced signal (more positive).
These trends indicate that cation solvation locally disrupts the orientational structure of interfacial water.
Turning to the Cl$^-$-solvating waters (Figure~\ref{fig:fig3}d), we see a milder effect.
The hydrogen-bonded region remains largely similar to that of pure water, with only a slight reduction in magnitude and no clear dependence on concentration.
In the dangling O--H region, there is a moderate increase in signal (smaller than that seen for Na$^+$) again with minimal variation across concentrations.

Taken together, these results reveal the origin of the overall spectral changes: the slight weakening of the hydrogen-bonded O--H band primarily arises from Na$^+$ solvation, while the increase in dangling O--H signal reflects contributions from both cations and anions, with Na$^+$ again playing the dominant role.
Although the spectral response of Na$^{+}$-solvating water molecules differs markedly from that of interfacial water not solvating ions, their overall contribution to $\chi_{\textrm{total}}^{(2)}$ remains modest because such molecules represent only a minority of the interfacial population.
As a result, the total spectrum reflects a weighted sum dominated by unperturbed water molecules (Figure~\ref{fig:fig3}a).

\subsection*{Ion-Specific Coordination Governs Interfacial Water Orientation}

We now discuss the origins of the subtle spectral changes.
To make these effects clearer, we focus on the 2~M NaCl case, where the differences are most pronounced.
We begin by looking at the water molecules that do not solvate ions.
Figure~\ref{fig:fig4}a shows the two-dimensional (2D) probability distribution of O--H bond orientations as a function of depth from the graphene interface.
Each water molecule contributes two O--H bond vectors; an angle of $0^\circ$ corresponds to O--H bonds pointing into the bulk, whereas $180^\circ$ corresponds to bonds pointing toward the interface.
The distribution shows a dominant feature around $100^\circ$, corresponding to O--H bonds lying largely parallel to the surface, and a weaker feature near $25^\circ$, associated with a small population of bonds pointing into the bulk.
The corresponding 1D projection, shown on the right side of the 2D map, highlights these two features more clearly, with peaks around $25^\circ$ and $100^\circ$, consistent with previous observations of water near hydrophobic surfaces~\cite{Ruiz-Barragan_JPCL_2019, Dufils_ChemSci_2024}.
Most interfacial water molecules therefore adopt in-plane orientations that participate in an extended two-dimensional hydrogen-bond network, while only a minority point out of plane.
Because VSFG is sensitive primarily to the out-of-plane component of molecular orientations at the interface, we focus on these less abundant configurations, as they are the ones that determine the spectral differences observed under the particular polarization combination used here.
The overall orientation pattern, including both the in-plane network and the SFG-active out-of-plane bonds, is schematically illustrated in Figure~\ref{fig:fig4}c.
To assess the influence of ions, we compute the difference between this distribution and that for pure water.
The difference map confirms that the orientation pattern of water molecules not directly coordinating ions remains largely unchanged (Figure~\ref{fig:fig4}b).
Only minor deviations are observed, with negative regions indicating weakened features and positive regions indicating emerging ones.
Thus, we detect no significant rearrangement beyond the first solvation shell, suggesting that ion effects do not propagate to non-coordinating interfacial water molecules, unlike the longer-range perturbations observed in bulk water~\cite{tielrooij_2010}.

We now turn to the analysis of water molecules solvating Na$^{+}$.
As shown in the 2D probability distribution of O--H bond orientations in Figure~\ref{fig:fig4}d, a single dominant maximum emerges.
Comparing this distribution to that of pure water (Figure~\ref{fig:fig4}e), we find a pronounced depletion of orientations near $25^\circ$, corresponding to the loss of the downward-pointing interfacial configuration.
In contrast, there is a modest increase in the probability of larger angles (above $100^\circ$). 
This reduction in hydrogen-bonded O--H orientations, together with the enhanced occurrence of dangling O--H bonds, is consistent with the decrease in the hydrogen-bonded O--H band and the increase in the dangling O--H peak observed in the VSFG spectra (Figure~\ref{fig:fig3}c).
A schematic representation of water orientations around Na$^{+}$ at the interface is provided in Figure~\ref{fig:fig4}f.
Importantly, Na$^{+}$ resides slightly below the average interfacial plane defined by the oxygen atoms of the first water layer (Figure~\ref{fig:fig2}a).
This offset in depth, together with the characteristic dipolar alignment for cations, rationalizes both the observed angular preferences and the corresponding changes in the spectral response.

Lastly, we examine water molecules that solvate Cl$^{-}$.
As shown in the 2D probability distribution in Figure~\ref{fig:fig4}g, the orientational profile closely resembles that of water molecules not solvating ions, consistent with the minimal spectral changes observed in the VSFG spectra discussed earlier.
The corresponding difference map with respect to pure water (Figure~\ref{fig:fig4}h) reveals only subtle changes, primarily a depletion of O--H bond orientations around $25^\circ$. 
A schematic of the molecular orientations is shown in Figure~\ref{fig:fig4}i.
In contrast to the dipolar alignment seen for Na$^{+}$, Cl$^{-}$ solvates water via hydrogen bonding, with one of the hydrogen atoms pointing directly toward the anion.
The overall orientation remains broadly similar to that of water molecules not solvating ions, except for this additional hydrogen-bond donor interaction.
This O--H bond typically lies nearly parallel to the interface, making it weakly VSFG-active because it lacks a significant out-of-plane projection.
Consequently, Cl$^{-}$ ions primarily perturb O–H bonds that are spectroscopically inactive, which explains the relatively modest spectral changes associated with Cl$^{-}$ solvation.

For completeness, we also analyzed the microscopic origins of the subtle variations observed in the free O–H region.
This analysis, presented in detail in Section~S3, confirms that these weaker spectral features primarily arise from fine adjustments in the orientations of interfacial water molecules.

\subsection*{Ions Disrupt the Extended Interface Hydrogen-Bond Network}

\begin{figure}[htp!]
    \centering
    \includegraphics[width=0.45\textwidth]{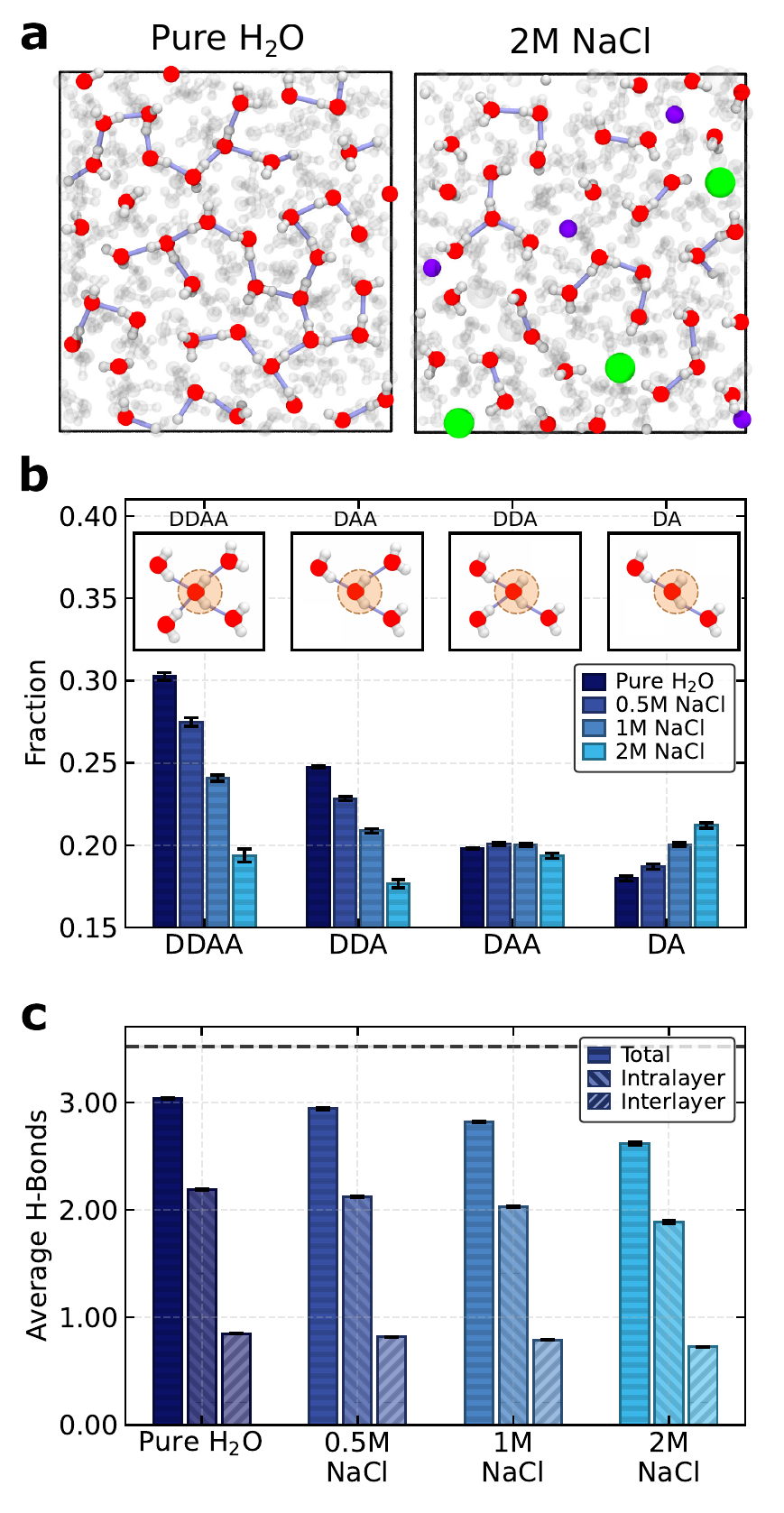}
    \caption{\textbf{Ion-induced reorganization of the interfacial hydrogen-bond network.}
    (a) Snapshots of the interfacial water structure at the graphene–water interface for pure water and a 2~M NaCl solution.
    Sodium ions are shown in purple, and chloride ions are in green.
    (b) Distribution of hydrogen-bond topologies at different NaCl concentrations.
    Fractions of DDAA, DDA, DAA, and DA motifs are reported, where D and A denote hydrogen-bond donors and acceptors, respectively.
    The accompanying snapshots illustrate representative examples of the hydrogen-bonding motifs discussed.
    (c) Average number of hydrogen bonds per interfacial water molecule, decomposed into total, intralayer (within the topmost layer), and interlayer (between the first and second layers).
    The horizontal dashed blue line marks the bulk-water value.
    }
    \label{fig:fig5}
\end{figure}

So far, we have focused our analysis on the local water structure, characterized by the orientation and hydrogen bonding strength of the O–H bonds.
We now take a broader view and examine the topology and connectivity of the interfacial hydrogen-bond network, as well as how it is influenced by the presence of ions.
In the absence of salt, interfacial water forms an extended, collective 2D network with hydrogen bonds oriented largely parallel to the surface, consistent with previous observations at the air–water interface~\cite{pezzotti_2017} and other weakly interacting (hydrophobic) interfaces~\cite{pezzotti_2021_finger}  (Figure~\ref{fig:fig5}a).
Upon the addition of salt, this extended network becomes disrupted, forming fewer rings and exhibiting a slight shift in its distribution toward smaller chains and rings (Figure~S13).
This restructuring is also reflected in the hydrogen-bond topology: the fraction of DDAA motifs (two donors and two acceptors) decreases with increasing salt concentration.
In contrast, DA motifs (one donor and one acceptor) become more prevalent (Figure~\ref{fig:fig5}b).
At the same time, the average number of hydrogen bonds per water molecule declines with salt concentration, with reductions observed in both intralayer and interlayer bonds (Figure~\ref{fig:fig5}c)\cite{feng_2022}.
Together, these trends reveal a progressive weakening of the extended hydrogen-bond network.
Importantly, this restructuring occurs without a significant change in the total number of interfacial water molecules, indicating that ion adsorption reorganizes rather than depletes the interfacial layer.
The hydrogen-bond network thus adapts to accommodate ions while preserving the overall density and connectivity of interfacial water.

The restructuring results in a distinct spectroscopic fingerprint, most apparent at low frequencies within the hydrogen-bonded region of the surface-specific H vibrational density of states (Figure~S14).
In the far-infrared (terahertz, THz) band, at wavenumbers below about 300~cm$^{-1}$, the response is dominated by collective intermolecular motions, including hydrogen-bond translations and librations, as well as ion–water cage modes~\cite{martina_2020, sylvie_2024, Pezzotti2025_thz_natrev}.
The presence of salt redistributes spectral weight within the librational band, depleting cage-like motions (below~80 cm$^{-1}$) while enhancing intermolecular H-bond translations and the librational band (around~400~cm$^{-1}$)~\cite{marx_2010}.
These collective low-frequency modes represent a promising target for future THz-SFG studies~\cite{kramer_vsfg_thz_2016,pezzotti_2021_finger}, which could directly test how ions reshape the interfacial hydrogen-bond network.

\section*{Discussion}

We investigated the organization of ions and the structure of interfacial water at a prototypical hydrophobic interface: the NaCl(aq)–graphene interface.
In this system, ions accumulate strongly at the interface, yet the associated changes in the local structure of interfacial water are relatively modest.
This example clearly reveals that the standard paradigm linking ion surface propensity to interfacial water disruption—established for the air–water interface—is not universally applicable to other hydrophobic interfaces and should not be applied uncritically when interpreting experimental results. 
These differences likely underlie the reported inverse ordering of some molecular ions at graphene interfaces~\cite{nature_edl_2022,Hua_JPCL_2011, gra_wat_acid_2025}.

Simulating VSFG spectra is significantly more demanding than simulating bulk spectroscopies, as the signal originates from only a few molecular layers and requires extensive sampling to account for the cancellation of bulk contributions~\cite{ssvcf_2015, morita_book, yair_sfg_2023}.
The use of MLPs was therefore instrumental in this work, enabling multi-nanosecond simulations necessary to resolve the subtle spectral differences underlying our findings—well beyond the reach of direct \textit{ab initio} molecular dynamics.
In addition to their computational efficiency, MLPs offer near first-principles accuracy, which was essential for reliably modeling the interfacial properties.
Indeed, classical electrostatics predict a universal repulsion of ions from hydrophobic aqueous interfaces~\cite{onsager_1934, pavel_2006}, and standard force fields fail to reproduce even the basic enrichment of ions at graphitic surfaces~\cite{Horinek_PRL_2007, Cole_JPCL_2011, kara_pairing_2024}, rendering them inadequate for this problem.
Crucial to the reliability of our results is the explicit inclusion of long-range electrostatic effects~\cite{Niblett_JCP_2021}.

Focusing on the microscopic origins of these effects, we find that cations perturb the interfacial water structure more strongly than anions, yet these perturbations remain largely confined to the first solvation shell~\cite{klein_2022}: waters not directly coordinating either ion retain their characteristic interfacial arrangement. 
This short-ranged character of ionic perturbations suggests an enhanced local dielectric screening at the graphene–water interface driven by the graphene polarization.
The same effect has been suggested from continuum models~\cite{kavokine_2022} and atomistic simulations~\cite{kara_pairing_2024} to weaken cation–anion attraction, thereby decreasing the stability of ion pairs at the interface.
This behavior stands in clear contrast to the longer-ranged ionic perturbations reported in bulk solutions~\cite{Chen_sciadv_2016,tielrooij_2010}.
We note, however, that these are interfacial effects and should not be mistaken for confinement effects when analyzing graphene-based nanodevices~\cite{yongkang_natcoms_2025, reactivity_2025_interfacial_or_conf}.

The charge and rigidity of the interface play pivotal roles. 
Near the point of zero charge, subtle effects such as those reported here may emerge, whereas at higher charge densities, electrostatic contributions might dominate and potentially overshadow local solvation effects~\cite{jassar_Arxiv_2025}. 
Geometrical constraints can also strongly influence interfacial ion solvation.
At the air–water interface, for example, ions couple strongly to capillary wave fluctuations, thereby altering them and influencing properties such as surface tension~\cite{petersen_2006}. 
In contrast, the rigidity of graphene suppresses such fluctuations, diminishing their impact on interfacial organization~\cite{scox_pnas_2017}, while at the oil–water interface, such fluctuations are enhanced~\cite{Devlin_pnas_2022}.
In this work, we have focused on neutral and flat graphene, which provides a valuable model for nanodevices.
Future investigations will extend this approach to charged and flexible surfaces, representing natural next steps, as well as to different ion types.
Such studies will help assess the extent to which broader frameworks, such as the Hofmeister series, apply to solid hydrophobic interfaces.

Our results demonstrate that solid–liquid interfaces can host substantial populations of interfacial ions.
At 2~M NaCl, roughly half of the interfacial water molecules (about 45\%) are solvated by ions, without significantly altering either the number of interfacial water molecules or their local structure.
This ability of carbon-based materials to accommodate dense interfacial ion populations while maintaining a largely unperturbed water structure is particularly beneficial in nanofluidic and electrochemical contexts, where performance depends sensitively on the local organization of nearby water~\cite{siria_giant_2013, robin_2023, capacit_2008, batt_2}.
At the same time, NaCl disrupts the extended two-dimensional hydrogen-bond network, breaking it into smaller domains and weakening the connectivity between adjacent water layers, rendering the interlayer region more hydrophobic~\cite{pezzotti_2021_finger}.
This modulation could, in turn, be exploited to tune the local acid-base chemistry~\cite{havenith_e_gold_2024} and thereby influence electrochemical reactions~\cite{govindarajan_2022}, as well as to enhance the solubility of CO$_2$ in graphene-based supercapacitors~\cite{coady_2025}.
More broadly, this duality of local robustness and global lability emerges as a general feature that could be exploited to advance applications in energy storage, conversion, catalysis, and sensing.

\section*{Methods}

\textbf{Machine Learning Potential.}
Molecular dynamics simulations were performed using a committee of eight Behler–Parrinello neural network potentials (NNPs) trained on revPBE-D3(0) reference data~\cite{revpbed3_1, revpbed3_2} developed in our previous work~\cite{kara_pairing_2024}.
This exchange–correlation functional has been shown to describe water–graphene~\cite{angelos_dft_water_2016, ondrej_revpbe_2017} and NaCl–water interactions reliably~\cite{acs_nano_kara_2025}, and the resulting NNP reproduces \textit{ab initio} structural and dynamical properties~\cite{kara_pairing_2024, acs_nano_kara_2025}.
The short-range interactions were described using atom-centered symmetry functions with a 12~Bohr cutoff, while long-range electrostatics were included via a fixed-charge Coulomb baseline (+1 for Na$^{+}$, –1 for Cl$^{-}$, 0 for C, and SPC/E charges for water).
The total energy was expressed as $E = E_{\mathrm{sr}} + E_{\mathrm{Coul}}$, where the NNP was trained on $E_{\mathrm{sr}}$ only.
Additional details are provided in Section S1.

\textbf{Molecular Dynamics Simulations.} 
All simulations were carried out in LAMMPS interfaced with n2p2~\cite{n2p2_1, n2p2_2}.
Each system consisted of an aqueous NaCl solution confined between two rigid graphene sheets forming a slit pore with lateral dimensions of 19.76~$\times$~21.39~Å$^2$ (see Table S1 for further details).
Periodic boundary conditions were applied in all directions, and a vacuum layer three times the slit height was included along $z$ to remove spurious interactions.
Electrostatics were computed using the PPPM method, and the Yeh–Berkowitz correction~\cite{yeh_berkowitz} was applied to account for the slab geometry.
After 25~ps of NVT equilibration, production simulations of 200~ps were performed in the NVE ensemble, averaged over 40 independent trajectories, totaling over 30~ns of sampling.
Temperature was maintained at 300~K using a Nosé–Hoover thermostat with a 0.5~fs time step.
Graphene atoms were fixed during production runs.

\textbf{Data Analysis.}
VSFG spectra were computed via the surface-specific velocity–velocity correlation function (ssVVCF) approach~\cite{ssvcf_2015}.
The analysis focused on the topmost interfacial water layer, which dominates the spectral response (Figure S4).
We verified that moderate variations in these cutoffs do not qualitatively affect our conclusions.
The hydrogen bonds are identified using the geometric definition provided in Ref.~\citenum{luzar_hydrogen-bond_1996}.
Interfacial water molecules are defined as those located between the graphene surface and the first minimum of the water oxygen density profile (see Figure~S1), corresponding to a distance of approximately 4.5~{\AA} from the surface.
Additional details are provided in Section~S1.

\textbf{Sample Preparation and HD-VSFG Measurement.}
The preparation of suspended graphene on the water surface followed procedures similar to those reported previously~\cite{xu_structure_2023,nature_edl_2022,wang_ange_2024}.
Details have been presented in Ref.~\citenum{wang_ange_2024}.
HD-SFG measurements were performed using a noncollinear setup driven by a Ti:Sapphire regenerative amplifier laser system (800 nm central wavelength, ~40 fs pulse width, 5 mJ pulse energy, 1 kHz repetition rate).
The configuration of the optical setup has been described previously~\cite{seki_real-time_2021,wang_ange_2024}.
For heterodyne detection, the local oscillator (LO) was generated by focusing the IR and visible beams onto a 200 nm-thick ZnO film deposited on a 1 mm-thick CaF$_2$ substrate, following established procedures~\cite{vanselous_extending_2016}.
The LO, IR, and visible beams were subsequently directed and refocused using pairs of off-axis parabolic mirrors to achieve spatial and temporal overlap at the graphene-water interface.
The incidence angles (in air) for the IR, visible, and LO beams were $50^\circ$, $61^\circ$, and $64^\circ$, respectively.
Measurements were carried out under the $ssp$ polarization combination ($s$-polarized SFG and visible beams, $p$-polarized IR) in a dry air atmosphere to minimize interference from water vapor.
Phase referencing was performed using z-cut quartz, and the sample height was checked with a displacement sensor (CL-3000, Keyence).

\begin{acknowledgement}
We thank Stephen J. Cox for insightful comments on the paper.
X.R.A., Y.W., M.B., A.M., and Y.L. acknowledge support from the European Union under the “n-AQUA” European Research Council project (Grant No. 101071937). 
K.D.F. acknowledges support from Schmidt Science Fellows, in partnership with the Rhodes Trust, and Trinity College, Cambridge.
Y.W. and M.B. are also grateful for the financial support from the MaxWater Initiative of the Max Planck Society. 
C.S. acknowledges financial support from the Deutsche Forschungsgemeinschaft (German Research Foundation) Project No. 500244608, as well as from the Royal Society Grant No. RGS/R2/242614.
This work used the ARCHER2 UK National Supercomputing Service via the UK’s HEC Materials Chemistry Consortium, funded by EPSRC (EP/F067496).
We also utilized resources from the Cambridge Service for Data Driven Discovery (CSD3), supported by EPSRC (EP/T022159/1) and DiRAC funding, with additional access through a University of Cambridge EPSRC Core Equipment Award (EP/X034712/1).
We also acknowledge EuroHPC Joint Undertaking for awarding the project ID EHPC-REG-2024R02-130 access to Leonardo at CINECA, Italy.

\end{acknowledgement}

\section*{Data Availability}
All data required to reproduce the findings of this work will be made openly available on GitHub upon acceptance of this manuscript.

\bibliography{references}

@article{Cole_JPCL_2011,
author = {Cole, Daniel J. and Ang, Priscilla K. and Loh, Kian Ping},
title = {Ion Adsorption at the Graphene/Electrolyte Interface},
journal = {The Journal of Physical Chemistry Letters},
volume = {2},
number = {14},
pages = {1799-1803},
year = {2011},
doi = {10.1021/jz200765z},

}

@article{Horinek_PRL_2007,
  title = {Specific Ion Adsorption at Hydrophobic Solid Surfaces},
  author = {Horinek, Dominik and Netz, Roland R.},
  journal = {Phys. Rev. Lett.},
  volume = {99},
  issue = {22},
  pages = {226104},
  numpages = {4},
  year = {2007},
  month = {Nov},
  publisher = {American Physical Society},
  doi = {10.1103/PhysRevLett.99.226104},
  url = {https://link.aps.org/doi/10.1103/PhysRevLett.99.226104}
}

@BOOK{Dannenberg_1997,
  TITLE = {An Introduction to Hydrogen Bonding},
  AUTHOR = {George A. Jeffrey},
  YEAR = {1997},
  PUBLISHER = {Oxford Univ. Press},
}

@Article{Sedlmeier2008,
author={Sedlmeier, Felix
and Janecek, Jiri
and Sendner, Christian
and Bocquet, Lyderic
and Netz, Roland R.
and Horinek, Dominik},
title={Water at polar and nonpolar solid walls (Review)},
journal={Biointerphases},
year={2008},
month={Sep},
day={01},
volume={3},
number={3},
pages={FC23-FC39},
issn={1559-4106},
doi={10.1116/1.2999559},
url={https://doi.org/10.1116/1.2999559}
}

@article{Willard_JCP_2014,
    author = {Willard, Adam P. and Chandler, David},
    title = {The molecular structure of the interface between water and a hydrophobic substrate is liquid-vapor like},
    journal = {The Journal of Chemical Physics},
    volume = {141},
    number = {18},
    pages = {18C519},
    year = {2014},
    month = {10},
    issn = {0021-9606},
    doi = {10.1063/1.4897249},
    url = {https://doi.org/10.1063/1.4897249},
    eprint = {https://pubs.aip.org/aip/jcp/article-pdf/doi/10.1063/1.4897249/15491940/18c519_1_online.pdf},
}

@article{
Du_Science_1994,
author = {Quan Du  and Eric Freysz  and Y. Ron Shen },
title = {Surface Vibrational Spectroscopic Studies of Hydrogen Bonding and Hydrophobicity},
journal = {Science},
volume = {264},
number = {5160},
pages = {826-828},
year = {1994},
doi = {10.1126/science.264.5160.826},
URL = {https://www.science.org/doi/abs/10.1126/science.264.5160.826},
eprint = {https://www.science.org/doi/pdf/10.1126/science.264.5160.826}}

@article{
Vasudevan_PNAS_2014,
author = {Vasudevan Venkateshwaran  and Srivathsan Vembanur  and Shekhar Garde },
title = {Water-mediated ion–ion interactions are enhanced at the water vapor–liquid interface},
journal = {Proceedings of the National Academy of Sciences},
volume = {111},
number = {24},
pages = {8729-8734},
year = {2014},
doi = {10.1073/pnas.1403294111},
URL = {https://www.pnas.org/doi/abs/10.1073/pnas.1403294111},
eprint = {https://www.pnas.org/doi/pdf/10.1073/pnas.1403294111}}

@article{
Tian_PNAS_2009,
author = {C. S. Tian  and Y. R. Shen },
title = {Structure and charging of hydrophobic material/water interfaces studied by phase-sensitive sum-frequency vibrational spectroscopy},
journal = {Proceedings of the National Academy of Sciences},
volume = {106},
number = {36},
pages = {15148-15153},
year = {2009},
doi = {10.1073/pnas.0901480106},
URL = {https://www.pnas.org/doi/abs/10.1073/pnas.0901480106},
eprint = {https://www.pnas.org/doi/pdf/10.1073/pnas.0901480106}}

@Article{Chandler2005,
author={Chandler, David},
title={Interfaces and the driving force of hydrophobic assembly},
journal={Nature},
year={2005},
month={Sep},
day={01},
volume={437},
number={7059},
pages={640-647},
issn={1476-4687},
doi={10.1038/nature04162},
url={https://doi.org/10.1038/nature04162}
}

@article{BHUSHAN20111,
title = {Natural and biomimetic artificial surfaces for superhydrophobicity, self-cleaning, low adhesion, and drag reduction},
journal = {Progress in Materials Science},
volume = {56},
number = {1},
pages = {1-108},
year = {2011},
issn = {0079-6425},
doi = {https://doi.org/10.1016/j.pmatsci.2010.04.003},
url = {https://www.sciencedirect.com/science/article/pii/S0079642510000289},
author = {Bharat Bhushan and Yong Chae Jung}
}

@article{seki_real-time_2021,
title = {Real-time study of on-water chemistry: Surfactant monolayer-assisted growth of a crystalline quasi-2D polymer},
journal = {Chem},
volume = {7},
number = {10},
pages = {2758-2770},
year = {2021},
issn = {2451-9294},
doi = {https://doi.org/10.1016/j.chempr.2021.07.016},
url = {https://www.sciencedirect.com/science/article/pii/S2451929421003727},
author = {Takakazu Seki and Xiaoqing Yu and Peng Zhang and Chun-Chieh Yu and Kejun Liu and Lucas Gunkel and Renhao Dong and Yuki Nagata and Xinliang Feng and Mischa Bonn}
}

@article{xu_structure_2023,
	author = {Xu, Ying and Ma, You-Bo and Gu, Feng and Yang, Shan-Shan and Tian, Chuan-Shan},
	date = {2023/09/01},
	doi = {10.1038/s41586-023-06374-0},
	id = {Xu2023},
	isbn = {1476-4687},
	journal = {Nature},
	number = {7979},
	pages = {506--510},
	title = {Structure evolution at the gate-tunable suspended graphene--water interface},
	url = {https://doi.org/10.1038/s41586-023-06374-0},
	volume = {621},
	year = {2023}}

@article{Devlin_pnas_2022,
author = {Shane W. Devlin  and Ilan Benjamin  and Richard J. Saykally },
title = {On the mechanisms of ion adsorption to aqueous interfaces: air-water vs. oil-water},
journal = {Proceedings of the National Academy of Sciences},
volume = {119},
number = {42},
pages = {e2210857119},
year = {2022},
doi = {10.1073/pnas.2210857119},
URL = {https://www.pnas.org/doi/abs/10.1073/pnas.2210857119},
eprint = {https://www.pnas.org/doi/pdf/10.1073/pnas.2210857119}}

@Article{jassar_Arxiv_2025,
      title={Resist the surface field: the H-bond network decides if water aligns at metal electrodes}, 
      author={Mohammed Bin Jassar and Wei-tao Liu and Simone Pezzotti},
      year={2025},
      eprint={2506.18467},
      archivePrefix={arXiv},
      note={arxiv (Chemical Physics). Submission Date: 23 June 2025. URL: https://arxiv.org/abs/2506.18467 (accessed 2025-11-13)},
      primaryClass={physics.chem-ph} 
}

@Article{Pezzotti2025_thz_natrev,
author={Pezzotti, Simone
and Chen, Wanlin
and Novelli, Fabio
and Yu, Xiaoqing
and Hoberg, Claudius
and Havenith, Martina},
title={Terahertz calorimetry spotlights the role of water in biological processes},
journal={Nature Reviews Chemistry},
year={2025},
month={Jul},
day={01},
volume={9},
number={7},
pages={481-494},
issn={2397-3358},
doi={10.1038/s41570-025-00712-8},
url={https://doi.org/10.1038/s41570-025-00712-8}
}

@article{
Chen_sciadv_2016,
author = {Yixing Chen  and Halil I. Okur  and Nikolaos Gomopoulos  and Carlos Macias-Romero  and Paul S. Cremer  and Poul B. Petersen  and Gabriele Tocci  and David M. Wilkins  and Chungwen Liang  and Michele Ceriotti  and Sylvie Roke },
title = {Electrolytes induce long-range orientational order and free energy changes in the H-bond network of bulk water},
journal = {Science Advances},
volume = {2},
number = {4},
pages = {e1501891},
year = {2016},
doi = {10.1126/sciadv.1501891},
URL = {https://www.science.org/doi/abs/10.1126/sciadv.1501891},
eprint = {https://www.science.org/doi/pdf/10.1126/sciadv.1501891}}

@article{Niblett_JCP_2021,
    author = {Niblett, Samuel P. and Galib, Mirza and Limmer, David T.},
    title = {Learning intermolecular forces at liquid–vapor interfaces},
    journal = {The Journal of Chemical Physics},
    volume = {155},
    number = {16},
    pages = {164101},
    year = {2021},
    month = {10},
    issn = {0021-9606},
    doi = {10.1063/5.0067565},
}

@article{Hua_JPCL_2011,
author = {Hua, Wei and Jubb, Aaron M. and Allen, Heather C.},
title = {Electric Field Reversal of Na2SO4, (NH4)2SO4, and Na2CO3 Relative to CaCl2 and NaCl at the Air/Aqueous Interface Revealed by Heterodyne Detected Phase-Sensitive Sum Frequency},
journal = {The Journal of Physical Chemistry Letters},
volume = {2},
number = {20},
pages = {2515-2520},
year = {2011},
doi = {10.1021/jz200888t},

}

@article{kara_pairing_2024,
   author = {Kara D Fong and Barbara Sumić and Niamh O’Neill and Christoph Schran and Clare P Grey and Angelos Michaelides},
   doi = {10.1021/acs.nanolett.4c00890},
   issn = {1530-6984},
   issue = {16},
   journal = {Nano Letters},
   month = {4},
   pages = {5024-5030},
   publisher = {American Chemical Society},
   title = {The Interplay of Solvation and Polarization Effects on Ion Pairing in Nanoconfined Electrolytes},
   volume = {24},
   year = {2024},
}

@article{ssvcf_2015,
    author = {Ohto, Tatsuhiko and Usui, Kota and Hasegawa, Taisuke and Bonn, Mischa and Nagata, Yuki},
    title = {Toward ab initio molecular dynamics modeling for sum-frequency generation spectra; an efficient algorithm based on surface-specific velocity-velocity correlation function},
    journal = {The Journal of Chemical Physics},
    volume = {143},
    number = {12},
    pages = {124702},
    year = {2015},
    month = {09},
    issn = {0021-9606},
    doi = {10.1063/1.4931106},
    url = {https://doi.org/10.1063/1.4931106},
    eprint = {https://pubs.aip.org/aip/jcp/article-pdf/doi/10.1063/1.4931106/14799947/124702\_1\_online.pdf},
}

@article{koelsch_2007,
title = {Specific ion effects in physicochemical and biological systems: Simulations, theory and experiments},
journal = {Colloids and Surfaces A: Physicochemical and Engineering Aspects},
volume = {303},
number = {1},
pages = {110-136},
year = {2007},
issn = {0927-7757},
doi = {https://doi.org/10.1016/j.colsurfa.2007.03.040},
url = {https://www.sciencedirect.com/science/article/pii/S0927775707002610},
author = {P. Koelsch and P. Viswanath and H. Motschmann and V.L. Shapovalov and G. Brezesinski and H. Möhwald and Dominik Horinek and Roland R. Netz and K. Giewekemeyer and T. Salditt and H. Schollmeyer and Regine {von Klitzing} and Jean Daillant and Patrick Guenoun}
}

@article{cui_protein_hydro_2014,
author = {Cui, Di and Ou, Shuching and Peters, Eric and Patel, Sandeep},
title = {Ion-Specific Induced Fluctuations and Free Energetics of Aqueous Protein Hydrophobic Interfaces: Toward Connecting to Specific-Ion Behaviors at Aqueous Liquid–Vapor Interfaces},
journal = {The Journal of Physical Chemistry B},
volume = {118},
number = {17},
pages = {4490-4504},
year = {2014},
doi = {10.1021/jp4105294},
URL = { 
    
        https://doi.org/10.1021/jp4105294
    
    

},
eprint = { 
    
        https://doi.org/10.1021/jp4105294
    
    

}

}

@article{litman_surface_2024,
    title = {Surface stratification determines the interfacial water structure of simple electrolyte solutions},
    volume = {16},
    issn = {1755-4349},
    url = {https://doi.org/10.1038/s41557-023-01416-6},
    doi = {10.1038/s41557-023-01416-6},
    number = {4},
    journal = {Nature Chemistry},
    author = {Litman, Yair and Chiang, Kuo-Yang and Seki, Takakazu and Nagata, Yuki and Bonn, Mischa},
    month = apr,
    year = {2024},
    pages = {644--650},
}

@article{Ruiz-Barragan_JPCL_2019,
author = {Ruiz-Barragan, Sergi and Muñoz-Santiburcio, Daniel and Marx, Dominik},
title = {Nanoconfined Water within Graphene Slit Pores Adopts Distinct Confinement-Dependent Regimes},
journal = {The Journal of Physical Chemistry Letters},
volume = {10},
number = {3},
pages = {329-334},
year = {2019},
doi = {10.1021/acs.jpclett.8b03530},

}

@Article{Dufils_ChemSci_2024,
author ="Dufils, T. and Schran, C. and Chen, J. and Geim, A. K. and Fumagalli, L. and Michaelides, A.",
title  ="Origin of dielectric polarization suppression in confined water from first principles",
journal  ="Chem. Sci.",
year  ="2024",
volume  ="15",
issue  ="2",
pages  ="516-527",
publisher  ="The Royal Society of Chemistry",
doi  ="10.1039/D3SC04740G",
url  ="http://dx.doi.org/10.1039/D3SC04740G"}

@article{yuki_free_oh_2018,
author = {Tang, Fujie and Ohto, Tatsuhiko and Hasegawa, Taisuke and Xie, Wen Jun and Xu, Limei and Bonn, Mischa and Nagata, Yuki},
title = {Definition of Free O–H Groups of Water at the Air–Water Interface},
journal = {Journal of Chemical Theory and Computation},
volume = {14},
number = {1},
pages = {357-364},
year = {2018},
doi = {10.1021/acs.jctc.7b00566},

URL = { 
    
        https://doi.org/10.1021/acs.jctc.7b00566
    
    

},
eprint = { 
    
        https://doi.org/10.1021/acs.jctc.7b00566
    
    

}

}

@article{revpbed3_2,
   author = {Stefan Grimme and Jens Antony and Stephan Ehrlich and Helge Krieg},
   doi = {10.1063/1.3382344},
   issn = {0021-9606},
   issue = {15},
   journal = {The Journal of Chemical Physics},
   month = {4},
   pages = {154104},
   title = {A consistent and accurate ab initio parametrization of density functional dispersion correction (DFT-D) for the 94 elements H-Pu},
   volume = {132},
   url = {https://doi.org/10.1063/1.3382344},
   year = {2010},
}

@article{revpbed3_1,
   author = {John P Perdew and Kieron Burke and Matthias Ernzerhof},
   doi = {10.1103/PhysRevLett.77.3865},
   issue = {18},
   journal = {Physical Review Letters},
   month = {10},
   pages = {3865-3868},
   publisher = {American Physical Society},
   title = {Generalized Gradient Approximation Made Simple},
   volume = {77},
   url = {https://link.aps.org/doi/10.1103/PhysRevLett.77.3865},
   year = {1996},
}

@article{acs_nano_kara_2025,
author = {Fong, Kara D. and Grey, Clare P. and Michaelides, Angelos},
title = {On the Physical Origins of Reduced Ionic Conductivity in Nanoconfined Electrolytes},
journal = {ACS Nano},
volume = {19},
number = {13},
pages = {13191-13201},
year = {2025},
doi = {10.1021/acsnano.4c18956},

URL = { 
    
        https://doi.org/10.1021/acsnano.4c18956
    
    

},
eprint = { 
    
        https://doi.org/10.1021/acsnano.4c18956
    
    

}

}

@article{n2p2_1,
author = {Singraber, Andreas and Behler, J{\"o}rg and Dellago, Christoph},
title = {Library-Based LAMMPS Implementation of High-Dimensional Neural Network Potentials},
journal = {Journal of Chemical Theory and Computation},
volume = {15},
number = {3},
pages = {1827-1840},
year = {2019},
doi = {10.1021/acs.jctc.8b00770},
URL = { 
    
        https://doi.org/10.1021/acs.jctc.8b00770
    
    

},
eprint = { 
        https://doi.org/10.1021/acs.jctc.8b00770
}
}

@article{dang_params,
author = {Dang, Liem X.},
title = {Mechanism and Thermodynamics of Ion Selectivity in Aqueous Solutions of 18-Crown-6 Ether: A Molecular Dynamics Study},
journal = {Journal of the American Chemical Society},
volume = {117},
number = {26},
pages = {6954-6960},
year = {1995},
doi = {10.1021/ja00131a018},

URL = { 
    
        https://doi.org/10.1021/ja00131a018
    
    

},
eprint = { 
    
        https://doi.org/10.1021/ja00131a018
    
    

}

}

@article{gra_wat_acid_2025,
author = {Advincula, Xavier R. and Fong, Kara D. and Michaelides, Angelos and Schran, Christoph},
title = {Protons Accumulate at the Graphene–Water Interface},
journal = {ACS Nano},
volume = {19},
number = {18},
pages = {17728-17737},
year = {2025},
doi = {10.1021/acsnano.5c02053},

URL = { 
    
        https://doi.org/10.1021/acsnano.5c02053
    
    

},
eprint = { 
    
        https://doi.org/10.1021/acsnano.5c02053
    
    

}

}

@article{atmos_1,
author = {K. W. Oum  and M. J. Lakin  and D. O. DeHaan  and T. Brauers  and B. J. Finlayson-Pitts },
title = {Formation of Molecular Chlorine from the Photolysis of Ozone and Aqueous Sea-Salt Particles},
journal = {Science},
volume = {279},
number = {5347},
pages = {74-76},
year = {1998},
doi = {10.1126/science.279.5347.74},
URL = {https://www.science.org/doi/abs/10.1126/science.279.5347.74},
eprint = {https://www.science.org/doi/pdf/10.1126/science.279.5347.74}}

@article{clare_2016,
author = {Forse, Alexander C. and Merlet, C{\'e}line and Griffin, John M. and Grey, Clare P.},
title = {New Perspectives on the Charging Mechanisms of Supercapacitors},
journal = {Journal of the American Chemical Society},
volume = {138},
number = {18},
pages = {5731-5744},
year = {2016},
doi = {10.1021/jacs.6b02115},

URL = { 
    
        https://doi.org/10.1021/jacs.6b02115
    
    

},
eprint = { 
    
        https://doi.org/10.1021/jacs.6b02115
    
    

}

}

@article{heather_2014,
author = {Hua, Wei and Verreault, Dominique and Huang, Zishuai and Adams, Ellen M. and Allen, Heather C.},
title = {Cation Effects on Interfacial Water Organization of Aqueous Chloride Solutions. I. Monovalent Cations: Li+, Na+, K+, and NH4+},
journal = {The Journal of Physical Chemistry B},
volume = {118},
number = {28},
pages = {8433-8440},
year = {2014},
doi = {10.1021/jp503132m},

URL = { 
    
        https://doi.org/10.1021/jp503132m
    
    

},
eprint = { 
    
        https://doi.org/10.1021/jp503132m
    
    

}

}

@article{shen_2011,
author = {Tian, Chuanshan and Byrnes, Steven J. and Han, Hui-Ling and Shen, Y. Ron},
title = {Surface Propensities of Atmospherically Relevant Ions in Salt Solutions Revealed by Phase-Sensitive Sum Frequency Vibrational Spectroscopy},
journal = {The Journal of Physical Chemistry Letters},
volume = {2},
number = {15},
pages = {1946-1949},
year = {2011},
doi = {10.1021/jz200791c},

URL = { 
    
        https://doi.org/10.1021/jz200791c
    
    

},
eprint = { 
    
        https://doi.org/10.1021/jz200791c
    
    

}

}

@article{shen_og_1989_nature,
	author = {Shen, Y.  R. },
	date = {1989/02/01},
	doi = {10.1038/337519a0},
	id = {Shen1989},
	isbn = {1476-4687},
	journal = {Nature},
	number = {6207},
	pages = {519--525},
	title = {Surface properties probed by second-harmonic and sum-frequency generation},
	url = {https://doi.org/10.1038/337519a0},
	volume = {337},
	year = {1989}}

@article{shen_2016_prl,
  title = {Unveiling Microscopic Structures of Charged Water Interfaces by Surface-Specific Vibrational Spectroscopy},
  author = {Wen, Yu-Chieh and Zha, Shuai and Liu, Xing and Yang, Shanshan and Guo, Pan and Shi, Guosheng and Fang, Haiping and Shen, Y. Ron and Tian, Chuanshan},
  journal = {Phys. Rev. Lett.},
  volume = {116},
  issue = {1},
  pages = {016101},
  numpages = {5},
  year = {2016},
  month = {Jan},
  publisher = {American Physical Society},
  doi = {10.1103/PhysRevLett.116.016101},
  url = {https://link.aps.org/doi/10.1103/PhysRevLett.116.016101}
}

@article{Bonn_ANIE_2015_review,
author = {Bonn, Mischa and Nagata, Yuki and Backus, Ellen H. G.},
title = {Molecular Structure and Dynamics of Water at the Water–Air Interface Studied with Surface-Specific Vibrational Spectroscopy},
journal = {Angewandte Chemie International Edition},
volume = {54},
number = {19},
pages = {5560-5576},
doi = {https://doi.org/10.1002/anie.201411188},
url = {https://onlinelibrary.wiley.com/doi/abs/10.1002/anie.201411188},
eprint = {https://onlinelibrary.wiley.com/doi/pdf/10.1002/anie.201411188},
year = {2015}
}

@article{bonn_science_2014,
author = {Dan Lis  and Ellen H. G. Backus  and Johannes Hunger  and Sapun H. Parekh  and Mischa Bonn },
title = {Liquid flow along a solid surface reversibly alters interfacial chemistry},
journal = {Science},
volume = {344},
number = {6188},
pages = {1138-1142},
year = {2014},
doi = {10.1126/science.1253793},
URL = {https://www.science.org/doi/abs/10.1126/science.1253793},
eprint = {https://www.science.org/doi/pdf/10.1126/science.1253793}}

@article{shen_2008,
author = {Tian, Chuanshan and Ji, Na and Waychunas, Glenn A. and Shen, Y. Ron},
title = {Interfacial Structures of Acidic and Basic Aqueous Solutions},
journal = {Journal of the American Chemical Society},
volume = {130},
number = {39},
pages = {13033-13039},
year = {2008},
doi = {10.1021/ja8021297},

URL = { 
    
        https://doi.org/10.1021/ja8021297
    
    

},
eprint = { 
    
        https://doi.org/10.1021/ja8021297
    
    

}

}

@article{wang_ange_2024,
   author = {Yongkang Wang and Fujie Tang and Xiaoqing Yu and Tatsuhiko Ohto and Yuki Nagata and Mischa Bonn},
   doi = {https://doi.org/10.1002/anie.202319503},
   issn = {1433-7851},
   issue = {n/a},
   journal = {Angewandte Chemie International Edition},
   month = {3},
   pages = {e202319503},
   publisher = {John Wiley & Sons, Ltd},
   title = {Heterodyne-Detected Sum-Frequency Generation Vibrational Spectroscopy Reveals Aqueous Molecular Structure at the Suspended Graphene/Water Interface},
   volume = {n/a},
   url = {https://doi.org/10.1002/anie.202319503},
   year = {2024},
}

@article{scalfi_2024,
    author = {Scalfi, Laura and Lehmann, Louis and dos Santos, Alexandre P. and Becker, Maximilian R. and Netz, Roland R.},
    title = "{Propensity of hydroxide and hydronium ions for the air–water and graphene–water interfaces from ab initio and force field simulations}",
    journal = {The Journal of Chemical Physics},
    volume = {161},
    number = {14},
    pages = {144701},
    year = {2024},
    month = {10},
    issn = {0021-9606},
    doi = {10.1063/5.0226966},
    url = {https://doi.org/10.1063/5.0226966}
}

@article{pezzotti_2021_finger,
author = {Pezzotti, Simone and Serva, Alessandra and Sebastiani, Federico and Brigiano, Flavio Siro and Galimberti, Daria Ruth and Potier, Louis and Alfarano, Serena and Schwaab, Gerhard and Havenith, Martina and Gaigeot, Marie-Pierre},
title = {Molecular Fingerprints of Hydrophobicity at Aqueous Interfaces from Theory and Vibrational Spectroscopies},
journal = {The Journal of Physical Chemistry Letters},
volume = {12},
number = {15},
pages = {3827-3836},
year = {2021},
doi = {10.1021/acs.jpclett.1c00257},

URL = { 
    
        https://doi.org/10.1021/acs.jpclett.1c00257
    
    

},
eprint = { 
    
        https://doi.org/10.1021/acs.jpclett.1c00257
    
    

}

}

@article{pavel_2006,
author = {Jungwirth, Pavel and Tobias, Douglas J.},
title = {Specific Ion Effects at the Air/Water Interface},
journal = {Chemical Reviews},
volume = {106},
number = {4},
pages = {1259-1281},
year = {2006},
doi = {10.1021/cr0403741},

URL = { 
    
        https://doi.org/10.1021/cr0403741
    
    

},
eprint = { 
    
        https://doi.org/10.1021/cr0403741
    
    

}

}

@article{Shultz_2000,
author = {Mary Jane Shultz and Cheryl Schnitzer and Danielle Simonelli and Steve Baldelli},
title = {Sum frequency generation spectroscopy of the aqueous interface: Ionic and soluble molecular solutions},
journal = {International Reviews in Physical Chemistry},
volume = {19},
number = {1},
pages = {123--153},
year = {2000},
publisher = {Taylor \& Francis},
doi = {10.1080/014423500229882},


URL = { 
    
        https://doi.org/10.1080/014423500229882
    
    

},
eprint = { 
    
        https://doi.org/10.1080/014423500229882
    
    

}

}

@article{desalination_1,
author = {Menachem Elimelech  and William A. Phillip },
title = {The Future of Seawater Desalination: Energy, Technology, and the Environment},
journal = {Science},
volume = {333},
number = {6043},
pages = {712-717},
year = {2011},
doi = {10.1126/science.1200488},
URL = {https://www.science.org/doi/abs/10.1126/science.1200488},
eprint = {https://www.science.org/doi/pdf/10.1126/science.1200488}}

@article{forse_nanop_2024,
author = {Xinyu Liu  and Dongxun Lyu  and Céline Merlet  and Matthew J. A. Leesmith  and Xiao Hua  and Zhen Xu  and Clare P. Grey  and Alexander C. Forse },
title = {Structural disorder determines capacitance in nanoporous carbons},
journal = {Science},
volume = {384},
number = {6693},
pages = {321-325},
year = {2024},
doi = {10.1126/science.adn6242},
URL = {https://www.science.org/doi/abs/10.1126/science.adn6242},
eprint = {https://www.science.org/doi/pdf/10.1126/science.adn6242}}

@article{atmos_2,
author = {E. M. Knipping  and M. J. Lakin  and K. L. Foster  and P. Jungwirth  and D. J. Tobias  and R. B. Gerber  and D. Dabdub  and B. J. Finlayson-Pitts },
title = {Experiments and Simulations of Ion-Enhanced Interfacial Chemistry on Aqueous NaCl Aerosols},
journal = {Science},
volume = {288},
number = {5464},
pages = {301-306},
year = {2000},
doi = {10.1126/science.288.5464.301},
URL = {https://www.science.org/doi/abs/10.1126/science.288.5464.301},
eprint = {https://www.science.org/doi/pdf/10.1126/science.288.5464.301}}

@article{das_ez_2020,
   author = {Sudipta Das and Sho Imoto and Shumei Sun and Yuki Nagata and Ellen H G Backus and Mischa Bonn},
   issn = {0002-7863},
   issue = {2},
   journal = {Journal of the American Chemical Society},
   month = {1},
   pages = {945-952},
   publisher = {American Chemical Society},
   title = {Nature of Excess Hydrated Proton at the Water–Air Interface},
    doi = {10.1021/jacs.9b10807},
   volume = {142},
   year = {2020},
}

@article{sfg_proton_laage_2024,
author = {de la Puente, Miguel and Gomez, Axel and Laage, Damien},
title = {Neural Network-Based Sum-Frequency Generation Spectra of Pure and Acidified Water Interfaces with Air},
journal = {The Journal of Physical Chemistry Letters},
volume = {15},
number = {11},
pages = {3096-3102},
year = {2024},
doi = {10.1021/acs.jpclett.4c00113},

URL = { 
    
        https://doi.org/10.1021/acs.jpclett.4c00113
    
    

},
eprint = { 
    
        https://doi.org/10.1021/acs.jpclett.4c00113
    
    

}

}

@article{havenith_e_gold_2024,
author = {Murke, Steffen and Chen, Wanlin and Pezzotti, Simone and Havenith, Martina},
title = {Tuning Acid–Base Chemistry at an Electrified Gold/Water Interface},
journal = {Journal of the American Chemical Society},
volume = {146},
number = {18},
pages = {12423-12430},
year = {2024},
doi = {10.1021/jacs.3c13633},

URL = { 
    
        https://doi.org/10.1021/jacs.3c13633
    
    

},
eprint = { 
    
        https://doi.org/10.1021/jacs.3c13633
    
    

}

}

@article{coady_2025,
author = {Coady, Zeke and Brookes, Samuel G. H. and Shen, Zhaohan and Rhodes, Benjamin J. and Mapstone, Grace and Xu, Zhen and Yu, Wei and Nishihara, Hirotomo and Schran, Christoph and Michaelides, Angelos and Forse, Alexander C.},
title = {Unexpected Oversolubility of CO2 Measured at Electrode–Electrolyte Interfaces},
journal = {Journal of the American Chemical Society},
volume = {147},
number = {40},
pages = {36310-36319},
year = {2025},
doi = {10.1021/jacs.5c09712},

URL = { 
    
        https://doi.org/10.1021/jacs.5c09712
    
    

},
eprint = { 
    
        https://doi.org/10.1021/jacs.5c09712
    
    

}

}

@Article{kramer_vsfg_thz_2016,
author ="Tong, Yujin and Kampfrath, Tobias and Campen, R. Kramer",
title  ="Experimentally probing the libration of interfacial water: the rotational potential of water is stiffer at the air/water interface than in bulk liquid",
journal  ="Phys. Chem. Chem. Phys.",
year  ="2016",
volume  ="18",
issue  ="27",
pages  ="18424-18430",
publisher  ="The Royal Society of Chemistry",
doi  ="10.1039/C6CP01004K",
url  ="http://dx.doi.org/10.1039/C6CP01004K"}

@article{govindarajan_2022,
author = {Nitish Govindarajan  and Aoni Xu  and Karen Chan },
title = {How pH affects electrochemical processes},
journal = {Science},
volume = {375},
number = {6579},
pages = {379-380},
year = {2022},
doi = {10.1126/science.abj2421},
URL = {https://www.science.org/doi/abs/10.1126/science.abj2421},
eprint = {https://www.science.org/doi/pdf/10.1126/science.abj2421}}

@article{robin_2023,
author = {P. Robin  and T. Emmerich  and A. Ismail  and A. Niguès  and Y. You  and G.-H. Nam  and A. Keerthi  and A. Siria  and A. K. Geim  and B. Radha  and L. Bocquet },
title = {Long-term memory and synapse-like dynamics in two-dimensional nanofluidic channels},
journal = {Science},
volume = {379},
number = {6628},
pages = {161-167},
year = {2023},
doi = {10.1126/science.adc9931},
URL = {https://www.science.org/doi/abs/10.1126/science.adc9931},
eprint = {https://www.science.org/doi/pdf/10.1126/science.adc9931}}

@article{yongkang_natcoms_2025,
	author = {Wang, Yongkang and Tang, Fujie and Yu, Xiaoqing and Chiang, Kuo-Yang and Yu, Chun-Chieh and Ohto, Tatsuhiko and Chen, Yunfei and Nagata, Yuki and Bonn, Mischa},
	date = {2025/08/07},
	doi = {10.1038/s41467-025-62625-w},
	id = {Wang2025},
	isbn = {2041-1723},
	journal = {Nature Communications},
	number = {1},
	pages = {7288},
	title = {Interfaces govern the structure of angstrom-scale confined water solutions},
	url = {https://doi.org/10.1038/s41467-025-62625-w},
	volume = {16},
	year = {2025}}

@article{bonn_rev_2021,
	author = {Gonella, Grazia and Backus, Ellen H. G. and Nagata, Yuki and Bonthuis, Douwe J. and Loche, Philip and Schlaich, Alexander and Netz, Roland R. and K{\"u}hnle, Angelika and McCrum, Ian T. and Koper, Marc T. M. and Wolf, Martin and Winter, Bernd and Meijer, Gerard and Campen, R. Kramer and Bonn, Mischa},
	date = {2021/07/01},
	doi = {10.1038/s41570-021-00293-2},
	id = {Gonella2021},
	isbn = {2397-3358},
	journal = {Nature Reviews Chemistry},
	number = {7},
	pages = {466--485},
	title = {Water at charged interfaces},
	url = {https://doi.org/10.1038/s41570-021-00293-2},
	volume = {5},
	year = {2021}}

@article{nature_edl_2022,
author = {Yang, Shanshan and Zhao, Xiao and Lu, Yi-Hsien and Barnard, Edward S. and Yang, Peidong and Baskin, Artem and Lawson, John W. and Prendergast, David and Salmeron, Miquel},
title = {Nature of the Electrical Double Layer on Suspended Graphene Electrodes},
journal = {Journal of the American Chemical Society},
volume = {144},
number = {29},
pages = {13327-13333},
year = {2022},
doi = {10.1021/jacs.2c03344},

URL = { 
    
        https://doi.org/10.1021/jacs.2c03344
    
    

},
eprint = { 
    
        https://doi.org/10.1021/jacs.2c03344
    
    

}

}

@article{kavokine_2022,
    author = {Kavokine, Nikita and Robin, Paul and Bocquet, Lydéric},
    title = {Interaction confinement and electronic screening in two-dimensional nanofluidic channels},
    journal = {The Journal of Chemical Physics},
    volume = {157},
    number = {11},
    pages = {114703},
    year = {2022},
    month = {09},
    issn = {0021-9606},
    doi = {10.1063/5.0102002},
    url = {https://doi.org/10.1063/5.0102002},
    eprint = {https://pubs.aip.org/aip/jcp/article-pdf/doi/10.1063/5.0102002/16720454/114703_1_online.pdf},
}

@article{angelos_dft_water_2016,
   author = {Michael J Gillan and Dario Alfè and Angelos Michaelides},
   doi = {10.1063/1.4944633},
   issn = {0021-9606},
   issue = {13},
   journal = {The Journal of Chemical Physics},
   month = {4},
   pages = {130901},
   title = {Perspective: How good is DFT for water?},
   volume = {144},
   url = {https://doi.org/10.1063/1.4944633},
   year = {2016},
}

@article{capacit_2008,
author = {Stoller, Meryl D. and Park, Sungjin and Zhu, Yanwu and An, Jinho and Ruoff, Rodney S.},
title = {Graphene-Based Ultracapacitors},
journal = {Nano Letters},
volume = {8},
number = {10},
pages = {3498-3502},
year = {2008},
doi = {10.1021/nl802558y},

URL = { 
    
        https://doi.org/10.1021/nl802558y
    
    

},
eprint = { 
    
        https://doi.org/10.1021/nl802558y
    
    

}

}

@article{batt_2,
author = {Yoo, EunJoo and Kim, Jedeok and Hosono, Eiji and Zhou, Hao-shen and Kudo, Tetsuichi and Honma, Itaru},
title = {Large Reversible Li Storage of Graphene Nanosheet Families for Use in Rechargeable Lithium Ion Batteries},
journal = {Nano Letters},
volume = {8},
number = {8},
pages = {2277-2282},
year = {2008},
doi = {10.1021/nl800957b},

URL = { 
    
        https://doi.org/10.1021/nl800957b
    
    

},
eprint = { 
    
        https://doi.org/10.1021/nl800957b
    
    

}

}

@article{lifetime_oh,
    author = {Asbury, John B. and Steinel, Tobias and Kwak, Kyungwon and Corcelli, S. A. and Lawrence, C. P. and Skinner, J. L. and Fayer, M. D.},
    title = {Dynamics of water probed with vibrational echo correlation spectroscopy},
    journal = {The Journal of Chemical Physics},
    volume = {121},
    number = {24},
    pages = {12431-12446},
    year = {2004},
    month = {12},
    issn = {0021-9606},
    doi = {10.1063/1.1818107},
    url = {https://doi.org/10.1063/1.1818107},
    eprint = {https://pubs.aip.org/aip/jcp/article-pdf/121/24/12431/19158604/12431_1_online.pdf},
}

@Article{feng_2022,
author ="Feng, Yixuan and Fang, Hongwei and Gao, Yitian and Ni, Ke",
title  ="Hierarchical clustering analysis of hydrogen bond networks in aqueous solutions",
journal  ="Phys. Chem. Chem. Phys.",
year  ="2022",
volume  ="24",
issue  ="16",
pages  ="9707-9717",
publisher  ="The Royal Society of Chemistry",
doi  ="10.1039/D2CP00099G",
url  ="http://dx.doi.org/10.1039/D2CP00099G"}

@article{petersen_2006,
   author = "Petersen, Poul B. and Saykally, Richard J.",
   title = "On the Nature of Ions at the Liquid Water Surface",
   journal= "Annual Review of Physical Chemistry",
   year = "2006",
   volume = "57",
   number = "Volume 57, 2006",
   pages = "333-364",
   doi = "https://doi.org/10.1146/annurev.physchem.57.032905.104609",
   url = "https://www.annualreviews.org/content/journals/10.1146/annurev.physchem.57.032905.104609",
   publisher = "Annual Reviews",
   issn = "1545-1593",
   type = "Journal Article",
  }

@article{pezzotti_2017,
author = {Pezzotti, Simone and Galimberti, Daria Ruth and Gaigeot, Marie-Pierre},
title = {2D H-Bond Network as the Topmost Skin to the Air–Water Interface},
journal = {The Journal of Physical Chemistry Letters},
volume = {8},
number = {13},
pages = {3133-3141},
year = {2017},
doi = {10.1021/acs.jpclett.7b01257},

URL = { 
    
        https://doi.org/10.1021/acs.jpclett.7b01257
    
    

},
eprint = { 
    
        https://doi.org/10.1021/acs.jpclett.7b01257
    
    

}

}

@article{netz_2012_rev,
   author = "Netz, Roland R. and Horinek, Dominik",
   title = "Progress in Modeling of Ion Effects at the Vapor/Water Interface", 
   journal= "Annual Review of Physical Chemistry",
   year = "2012",
   volume = "63",
   number = "Volume 63, 2012",
   pages = "401-418",
   doi = "https://doi.org/10.1146/annurev-physchem-032511-143813",
   url = "https://www.annualreviews.org/content/journals/10.1146/annurev-physchem-032511-143813",
   publisher = "Annual Reviews",
   issn = "1545-1593",
   type = "Journal Article",
  }

@Article{martina_2020,
AUTHOR = {Novelli, Fabio and Guchhait, Biswajit and Havenith, Martina},
TITLE = {Towards Intense THz Spectroscopy on Water: Characterization of Optical Rectification by GaP, OH1, and DSTMS at OPA Wavelengths},
JOURNAL = {Materials},
VOLUME = {13},
YEAR = {2020},
NUMBER = {6},
ARTICLE-NUMBER = {1311},
URL = {https://www.mdpi.com/1996-1944/13/6/1311},
PubMedID = {32183131},
ISSN = {1996-1944},
DOI = {10.3390/ma13061311}
}

@article{sylvie_2024,
author = {Mischa Flór  and David M. Wilkins  and Miguel de la Puente  and Damien Laage  and Giuseppe Cassone  and Ali Hassanali  and Sylvie Roke },
title = {Dissecting the hydrogen bond network of water: Charge transfer and nuclear quantum effects},
journal = {Science},
volume = {386},
number = {6726},
pages = {eads4369},
year = {2024},
doi = {10.1126/science.ads4369},
URL = {https://www.science.org/doi/abs/10.1126/science.ads4369},
eprint = {https://www.science.org/doi/pdf/10.1126/science.ads4369}}

@article{desalination_2,
	author = {Surwade, Sumedh P. and Smirnov, Sergei N. and Vlassiouk, Ivan V. and Unocic, Raymond R. and Veith, Gabriel M. and Dai, Sheng and Mahurin, Shannon M.},
	date = {2015/05/01},
	doi = {10.1038/nnano.2015.37},
	id = {Surwade2015},
	isbn = {1748-3395},
	journal = {Nature Nanotechnology},
	number = {5},
	pages = {459--464},
	title = {Water desalination using nanoporous single-layer graphene},
	url = {https://doi.org/10.1038/nnano.2015.37},
	volume = {10},
	year = {2015}}

@article{onsager_1934,
    author = {Onsager, Lars and Samaras, Nicholas N. T.},
    title = {The Surface Tension of Debye‐Hückel Electrolytes},
    journal = {The Journal of Chemical Physics},
    volume = {2},
    number = {8},
    pages = {528-536},
    year = {1934},
    month = {08},
    issn = {0021-9606},
    doi = {10.1063/1.1749522},
    url = {https://doi.org/10.1063/1.1749522},
    eprint = {https://pubs.aip.org/aip/jcp/article-pdf/2/8/528/18788133/528_1_online.pdf},
}

@article{wilkins_2015,
    author = {Wilkins, David M. and Manolopoulos, David E. and Dang, Liem X.},
    title = {Nuclear quantum effects in water exchange around lithium and fluoride ions},
    journal = {The Journal of Chemical Physics},
    volume = {142},
    number = {6},
    pages = {064509},
    year = {2015},
    month = {02},
    issn = {0021-9606},
    doi = {10.1063/1.4907554},
    url = {https://doi.org/10.1063/1.4907554},
    eprint = {https://pubs.aip.org/aip/jcp/article-pdf/doi/10.1063/1.4907554/14035051/064509_1_online.pdf},
}

@article{wilkins_2017,
author = {Wilkins, David M. and Manolopoulos, David E. and Pipolo, Silvio and Laage, Damien and Hynes, James T.},
title = {Nuclear Quantum Effects in Water Reorientation and Hydrogen-Bond Dynamics},
journal = {The Journal of Physical Chemistry Letters},
volume = {8},
number = {12},
pages = {2602-2607},
year = {2017},
doi = {10.1021/acs.jpclett.7b00979},

URL = { 
    
        https://doi.org/10.1021/acs.jpclett.7b00979
    
    

},
eprint = { 
    
        https://doi.org/10.1021/acs.jpclett.7b00979
    
    

}

}

@article{non_2024,
author = {O’Neill, Niamh and Shi, Benjamin X. and Fong, Kara and Michaelides, Angelos and Schran, Christoph},
title = {To Pair or not to Pair? Machine-Learned Explicitly-Correlated Electronic Structure for NaCl in Water},
journal = {The Journal of Physical Chemistry Letters},
volume = {15},
number = {23},
pages = {6081-6091},
year = {2024},
doi = {10.1021/acs.jpclett.4c01030},

URL = { 
    
        https://doi.org/10.1021/acs.jpclett.4c01030
    
    

},
eprint = { 
    
        https://doi.org/10.1021/acs.jpclett.4c01030
    
    

}

}

@article{ondrej_revpbe_2017,
   author = {Ondrej Marsalek and Thomas E Markland},
   doi = {10.1021/acs.jpclett.7b00391},
   issue = {7},
   journal = {The Journal of Physical Chemistry Letters},
   month = {4},
   pages = {1545-1551},
   publisher = {American Chemical Society},
   title = {Quantum Dynamics and Spectroscopy of Ab Initio Liquid Water: The Interplay of Nuclear and Electronic Quantum Effects},
   volume = {8},
   url = {https://doi.org/10.1021/acs.jpclett.7b00391},
   year = {2017},
}

@article{fellows_2024,
author = {Fellows, Alexander P. and Duque, \'Alvaro D\'iaz and Balos, Vasileios and Lehmann, Louis and Netz, Roland R. and Wolf, Martin and Th\"amer, Martin},
title = {Sum-Frequency Generation Spectroscopy of Aqueous Interfaces: The Role of Depth and Its Impact on Spectral Interpretation},
journal = {The Journal of Physical Chemistry C},
volume = {128},
number = {49},
pages = {20733-20750},
year = {2024},
doi = {10.1021/acs.jpcc.4c06650},

URL = { 
    
        https://doi.org/10.1021/acs.jpcc.4c06650
    
    

},
eprint = { 
    
        https://doi.org/10.1021/acs.jpcc.4c06650
    
    

}

}

@article{tielrooij_2010,
author = {K. J. Tielrooij  and N. Garcia-Araez  and M. Bonn  and H. J. Bakker },
title = {Cooperativity in Ion Hydration},
journal = {Science},
volume = {328},
number = {5981},
pages = {1006-1009},
year = {2010},
doi = {10.1126/science.1183512},
URL = {https://www.science.org/doi/abs/10.1126/science.1183512},
eprint = {https://www.science.org/doi/pdf/10.1126/science.1183512}}

@Article{ohto_gra_hbn_2018,
author ="Ohto, Tatsuhiko and Tada, Hirokazu and Nagata, Yuki",
title  ="Structure and dynamics of water at water–graphene and water–hexagonal boron-nitride sheet interfaces revealed by ab initio sum-frequency generation spectroscopy",
journal  ="Phys. Chem. Chem. Phys.",
year  ="2018",
volume  ="20",
issue  ="18",
pages  ="12979-12985",
publisher  ="The Royal Society of Chemistry",
doi  ="10.1039/C8CP01351A",
url  ="http://dx.doi.org/10.1039/C8CP01351A"}

@article{air_wat_sfg_paesani_2016,
author = {Medders, Gregory R. and Paesani, Francesco},
title = {Dissecting the Molecular Structure of the Air/Water Interface from Quantum Simulations of the Sum-Frequency Generation Spectrum},
journal = {Journal of the American Chemical Society},
volume = {138},
number = {11},
pages = {3912-3919},
year = {2016},
doi = {10.1021/jacs.6b00893},

URL = { 
    
        https://doi.org/10.1021/jacs.6b00893
    
    

},
eprint = { 
    
        https://doi.org/10.1021/jacs.6b00893
    
    

}

}

@article{gra_wat_sfg_charged_paesani_2025,
author = {Rashmi, Richa and Balogun, Toheeb O. and Azom, Golam and Agnew, Henry and Kumar, Revati and Paesani, Francesco},
title = {Revealing the Water Structure at Neutral and Charged Graphene/Water Interfaces through Quantum Simulations of Sum Frequency Generation Spectra},
journal = {ACS Nano},
volume = {19},
number = {4},
pages = {4876-4886},
year = {2025},
doi = {10.1021/acsnano.4c16486},

URL = { 
    
        https://doi.org/10.1021/acsnano.4c16486
    
    

},
eprint = { 
    
        https://doi.org/10.1021/acsnano.4c16486
    
    

}

}

@article{scox_pnas_2017,
author = {Debra L. McCaffrey  and Son C. Nguyen  and Stephen J. Cox  and Horst Weller  and A. Paul Alivisatos  and Phillip L. Geissler  and Richard J. Saykally },
title = {Mechanism of ion adsorption to aqueous interfaces: Graphene/water vs. air/water},
journal = {Proceedings of the National Academy of Sciences},
volume = {114},
number = {51},
pages = {13369-13373},
year = {2017},
doi = {10.1073/pnas.1702760114},
URL = {https://www.pnas.org/doi/abs/10.1073/pnas.1702760114},
eprint = {https://www.pnas.org/doi/pdf/10.1073/pnas.1702760114}}

@Article{klein_2022,
author={Zhang, Chunyi
and Yue, Shuwen
and Panagiotopoulos, Athanassios Z.
and Klein, Michael L.
and Wu, Xifan},
title = {Dissolving salt is not equivalent to applying a pressure on water},
journal={Nature Communications},
year={2022},
month={Feb},
day={10},
volume={13},
number={1},
pages={822},
issn={2041-1723},
doi={10.1038/s41467-022-28538-8},
url={https://doi.org/10.1038/s41467-022-28538-8}
}

@article{marx_2010,
author = {Matthias Heyden  and Jian Sun  and Stefan Funkner  and Gerald Mathias  and Harald Forbert  and Martina Havenith  and Dominik Marx },
title = {Dissecting the THz spectrum of liquid water from first principles via correlations in time and space},
journal = {Proceedings of the National Academy of Sciences},
volume = {107},
number = {27},
pages = {12068-12073},
year = {2010},
doi = {10.1073/pnas.0914885107},
URL = {https://www.pnas.org/doi/abs/10.1073/pnas.0914885107},
eprint = {https://www.pnas.org/doi/pdf/10.1073/pnas.0914885107}}

@article{brooksi_2025,
author = {Samuel G. H. Brookes  and Venkat Kapil  and Angelos Michaelides  and Christoph Schran },
title = {CO$_2$ hydration at the air--water interface: A surface-mediated ``in-and-out'' mechanism},
journal = {Proceedings of the National Academy of Sciences},
volume = {122},
number = {34},
pages = {e2502684122},
year = {2025},
doi = {10.1073/pnas.2502684122},
URL = {https://www.pnas.org/doi/abs/10.1073/pnas.2502684122},
eprint = {https://www.pnas.org/doi/pdf/10.1073/pnas.2502684122}}

@article{chiang_2020,
author = {Chiang, Kuo-Yang and Dalstein, Laetitia and Wen, Yu-Chieh},
title = {Affinity of Hydrated Protons at Intrinsic Water/Vapor Interface Revealed by Ion-Induced Water Alignment},
journal = {The Journal of Physical Chemistry Letters},
volume = {11},
number = {3},
pages = {696-701},
year = {2020},
doi = {10.1021/acs.jpclett.9b03520},

URL = { 
    
        https://doi.org/10.1021/acs.jpclett.9b03520
    
    

},
eprint = { 
    
        https://doi.org/10.1021/acs.jpclett.9b03520
    
    

}

}

@article{yair_sfg_2023,
author = {Litman, Yair and Lan, Jinggang and Nagata, Yuki and Wilkins, David M.},
title = {Fully First-Principles Surface Spectroscopy with Machine Learning},
journal = {The Journal of Physical Chemistry Letters},
volume = {14},
number = {36},
pages = {8175-8182},
year = {2023},
doi = {10.1021/acs.jpclett.3c01989},
URL = { 
    
        https://doi.org/10.1021/acs.jpclett.3c01989
    
    

},
eprint = { 
    
        https://doi.org/10.1021/acs.jpclett.3c01989
    
    

}

}

@article{wc_2010,
author = {Willard, Adam P. and Chandler, David},
title = {Instantaneous Liquid Interfaces},
journal = {The Journal of Physical Chemistry B},
volume = {114},
number = {5},
pages = {1954-1958},
year = {2010},
doi = {10.1021/jp909219k},

URL = { 
    
        https://doi.org/10.1021/jp909219k
    
    

},
eprint = { 
    
        https://doi.org/10.1021/jp909219k
    
    

}

}

@BOOK{morita_book,
  TITLE = {Theory of Sum Frequency Generation Spectroscopy},
  AUTHOR = {Morita, Akihiro},
  YEAR = {2018},
  PUBLISHER = {Springer Singapore},
}

@article{werder_params,
author = {Werder, T. and Walther, J. H. and Jaffe, R. L. and Halicioglu, T. and Koumoutsakos, P.},
title = {On the Water-Carbon Interaction for Use in Molecular Dynamics Simulations of Graphite and Carbon Nanotubes},
journal = {The Journal of Physical Chemistry B},
volume = {107},
number = {6},
pages = {1345-1352},
year = {2003},
doi = {10.1021/jp0268112},

URL = { 
    
        https://doi.org/10.1021/jp0268112
    
    

},
eprint = { 
    
        https://doi.org/10.1021/jp0268112
    
    

}

}

@article{benderskii_2008,
author = {Stiopkin, Igor V. and Jayathilake, Himali D. and Bordenyuk, Andrey N. and Benderskii, Alexander V.},
title = {Heterodyne-Detected Vibrational Sum Frequency Generation Spectroscopy},
journal = {Journal of the American Chemical Society},
volume = {130},
number = {7},
pages = {2271-2275},
year = {2008},
doi = {10.1021/ja076708w},

URL = { 
    
        https://doi.org/10.1021/ja076708w
    
    

},
eprint = { 
    
        https://doi.org/10.1021/ja076708w
    
    

}

}

@article{
shen_science_1994,
author = {Quan Du  and Eric Freysz  and Y. Ron Shen },
title = {Surface Vibrational Spectroscopic Studies of Hydrogen Bonding and Hydrophobicity},
journal = {Science},
volume = {264},
number = {5160},
pages = {826-828},
year = {1994},
doi = {10.1126/science.264.5160.826},
URL = {https://www.science.org/doi/abs/10.1126/science.264.5160.826},
eprint = {https://www.science.org/doi/pdf/10.1126/science.264.5160.826}}

@article{tahara_2015,
    author = {Nihonyanagi, Satoshi and Kusaka, Ryoji and Inoue, Ken-ichi and Adhikari, Aniruddha and Yamaguchi, Shoichi and Tahara, Tahei},
    title = {Accurate determination of complex $\chi^{(2)}$ spectrum of the air/water interface},
    journal = {The Journal of Chemical Physics},
    volume = {143},
    number = {12},
    pages = {124707},
    year = {2015},
    month = {09},
    issn = {0021-9606},
    doi = {10.1063/1.4931485},
    url = {https://doi.org/10.1063/1.4931485},
    eprint = {https://pubs.aip.org/aip/jcp/article-pdf/doi/10.1063/1.4931485/14799823/124707_1_online.pdf},
}

@article{tahara_2013_rev,
   author = "Nihonyanagi, Satoshi and Mondal, Jahur A. and Yamaguchi, Shoichi and Tahara, Tahei",
   title = "Structure and Dynamics of Interfacial Water Studied by Heterodyne-Detected Vibrational Sum-Frequency Generation", 
   journal= "Annual Review of Physical Chemistry",
   year = "2013",
   volume = "64",
   number = "Volume 64, 2013",
   pages = "579-603",
   doi = "https://doi.org/10.1146/annurev-physchem-040412-110138",
   url = "https://www.annualreviews.org/content/journals/10.1146/annurev-physchem-040412-110138",
   publisher = "Annual Reviews",
   issn = "1545-1593",
   type = "Journal Article",
  }

@article{yeh_berkowitz,
    author = {Yeh, In-Chul and Berkowitz, Max L.},
    title = {Ewald summation for systems with slab geometry},
    journal = {The Journal of Chemical Physics},
    volume = {111},
    number = {7},
    pages = {3155-3162},
    year = {1999},
    month = {08},
    issn = {0021-9606},
    doi = {10.1063/1.479595},
    url = {https://doi.org/10.1063/1.479595},
    eprint = {https://pubs.aip.org/aip/jcp/article-pdf/111/7/3155/19041975/3155\_1\_online.pdf},
}

@article{n2p2_2,
title = {LAMMPS - a flexible simulation tool for particle-based materials modeling at the atomic, meso, and continuum scales},
journal = {Computer Physics Communications},
volume = {271},
pages = {108171},
year = {2022},
issn = {0010-4655},
doi = {https://doi.org/10.1016/j.cpc.2021.108171},
url = {https://www.sciencedirect.com/science/article/pii/S0010465521002836},
author = {Aidan P. Thompson and H. Metin Aktulga and Richard Berger and Dan S. Bolintineanu and W. Michael Brown and Paul S. Crozier and Pieter J. {in 't Veld} and Axel Kohlmeyer and Stan G. Moore and Trung Dac Nguyen and Ray Shan and Mark J. Stevens and Julien Tranchida and Christian Trott and Steven J. Plimpton}
}

@article{bonn_2013_free_oh,
author = {Cho-Shuen Hsieh  and R. Kramer Campen  and Masanari Okuno  and Ellen H. G. Backus  and Yuki Nagata  and Mischa Bonn },
title = {Mechanism of vibrational energy dissipation of free OH groups at the air–water interface},
journal = {Proceedings of the National Academy of Sciences},
volume = {110},
number = {47},
pages = {18780-18785},
year = {2013},
doi = {10.1073/pnas.1314770110},
URL = {https://www.pnas.org/doi/abs/10.1073/pnas.1314770110},
eprint = {https://www.pnas.org/doi/pdf/10.1073/pnas.1314770110}}

@article{tahara_2008_up_down,
    author = {Yamaguchi, Shoichi and Tahara, Tahei},
    title = {Heterodyne-detected electronic sum frequency generation: “Up” versus “down” alignment of interfacial molecules},
    journal = {The Journal of Chemical Physics},
    volume = {129},
    number = {10},
    pages = {101102},
    year = {2008},
    month = {09},
    issn = {0021-9606},
    doi = {10.1063/1.2981179},
    url = {https://doi.org/10.1063/1.2981179},
    eprint = {https://pubs.aip.org/aip/jcp/article-pdf/doi/10.1063/1.2981179/16706825/101102_1_online.pdf},
}

@article{saykally_sfg_2018,
author = {Mizuno, Hikaru and Rizzuto, Anthony M. and Saykally, Richard J.},
title = {Charge-Transfer-to-Solvent Spectrum of Thiocyanate at the Air/Water Interface Measured by Broadband Deep Ultraviolet Electronic Sum Frequency Generation Spectroscopy},
journal = {The Journal of Physical Chemistry Letters},
volume = {9},
number = {16},
pages = {4753-4757},
year = {2018},
doi = {10.1021/acs.jpclett.8b01966},

URL = { 
    
        https://doi.org/10.1021/acs.jpclett.8b01966
    
    

},
eprint = { 
    
        https://doi.org/10.1021/acs.jpclett.8b01966
    
    

}

}

@article{chiang_2022,
author = {Kuo-Yang Chiang  and Takakazu Seki  and Chun-Chieh Yu  and Tatsuhiko Ohto  and Johannes Hunger  and Mischa Bonn  and Yuki Nagata },
title = {The dielectric function profile across the water interface through surface-specific vibrational spectroscopy and simulations},
journal = {Proceedings of the National Academy of Sciences},
volume = {119},
number = {36},
pages = {e2204156119},
year = {2022},
doi = {10.1073/pnas.2204156119},
URL = {https://www.pnas.org/doi/abs/10.1073/pnas.2204156119},
eprint = {https://www.pnas.org/doi/pdf/10.1073/pnas.2204156119}}

@article{yuki_dft_free_oh_2019,
author = {Ohto, Tatsuhiko and Dodia, Mayank and Xu, Jianhang and Imoto, Sho and Tang, Fujie and Zysk, Frederik and K{\"u}hne, Thomas D. and Shigeta, Yasuteru and Bonn, Mischa and Wu, Xifan and Nagata, Yuki},
title = {Accessing the Accuracy of Density Functional Theory through Structure and Dynamics of the Water–Air Interface},
journal = {The Journal of Physical Chemistry Letters},
volume = {10},
number = {17},
pages = {4914-4919},
year = {2019},
doi = {10.1021/acs.jpclett.9b01983},

URL = { 
    
        https://doi.org/10.1021/acs.jpclett.9b01983
    
    

},
eprint = { 
    
        https://doi.org/10.1021/acs.jpclett.9b01983
    
    

}

}

@article{luzar_hydrogen-bond_1996,
    title = {Hydrogen-bond kinetics in liquid water},
    volume = {379},
    issn = {1476-4687},
    url = {https://doi.org/10.1038/379055a0},
    doi = {10.1038/379055a0},
    number = {6560},
    journal = {Nature},
    author = {Luzar, Alenka and Chandler, David},
    month = jan,
    year = {1996},
    pages = {55--57},
}

@article{angelos_tocci_2014,
author = {Tocci, Gabriele and Joly, Laurent and Michaelides, Angelos},
title = {Friction of Water on Graphene and Hexagonal Boron Nitride from Ab Initio Methods: Very Different Slippage Despite Very Similar Interface Structures},
journal = {Nano Letters},
volume = {14},
number = {12},
pages = {6872-6877},
year = {2014},
doi = {10.1021/nl502837d},
URL = { 
    
        https://doi.org/10.1021/nl502837d
    
    

},
eprint = { 
    
        https://doi.org/10.1021/nl502837d
    
    

}

}

@misc{paesani_halide_air_wat_2025,
      title={Many-Body Interactions Govern Halide Distribution at the Air/Water Interface}, 
      author={Henry Agnew and Saswata Dasgupta and Francesco Paesani},
      year={2025},
      archivePrefix={ChemRxiv},
}

@Article{reactivity_2025_interfacial_or_conf,
      title={How reactive is water at the nanoscale and how to control it?}, 
      author={Xavier R. Advincula and Yair Litman and Kara D. Fong and William C. Witt and Christoph Schran and Angelos Michaelides},
      year={2025},
      eprint={2508.13034},
      archivePrefix={arXiv},
      note={arxiv (Chemical Physics). Submission Date: 18 Aug 2025. URL: https://arxiv.org/abs/2508.13034 (accessed 2025-11-13)},
      primaryClass={physics.chem-ph} 
}

@article{siria_giant_2013,
    title = {Giant osmotic energy conversion measured in a single transmembrane boron nitride nanotube},
    volume = {494},
    issn = {1476-4687},
    url = {https://doi.org/10.1038/nature11876},
    doi = {10.1038/nature11876},
    number = {7438},
    journal = {Nature},
    author = {Siria, Alessandro and Poncharal, Philippe and Biance, Anne-Laure and Fulcrand, Rémy and Blase, Xavier and Purcell, Stephen T. and Bocquet, Lydéric},
    month = feb,
    year = {2013},
    pages = {455--458},
}

@article{vanselous_extending_2016,
    title = {Extending the {Capabilities} of {Heterodyne}-{Detected} {Sum}-{Frequency} {Generation} {Spectroscopy}: {Probing} {Any} {Interface} in {Any} {Polarization} {Combination}},
    volume = {120},
    issn = {1932-7447, 1932-7455},
    shorttitle = {Extending the {Capabilities} of {Heterodyne}-{Detected} {Sum}-{Frequency} {Generation} {Spectroscopy}},
    url = {https://pubs.acs.org/doi/10.1021/acs.jpcc.6b01252},
    doi = {10.1021/acs.jpcc.6b01252},
    number = {15},
    urldate = {2022-03-14},
    journal = {The Journal of Physical Chemistry C},
    author = {Vanselous, Heather and Petersen, Poul B.},
    month = apr,
    year = {2016},
    pages = {8175--8184},
}

\end{document}


\def\mytitle{Breaking the Air–Water Paradigm: Ion Behavior at Hydrophobic Solid–Water Interfaces}
\title{Supporting Information for: \mytitle}

\author{Xavier R. Advincula}
\affiliation{Yusuf Hamied Department of Chemistry, University of Cambridge, Lensfield Road, Cambridge, CB2 1EW, UK}
\affiliation{Cavendish Laboratory, Department of Physics, University of Cambridge, Cambridge, CB3 0HE, UK}
\affiliation{Lennard-Jones Centre, University of Cambridge, Trinity Ln, Cambridge, CB2 1TN, UK}

\author{Kara D. Fong}
\affiliation{Division of Chemistry and Chemical Engineering, California Institute of Technology, Pasadena CA 91125, USA}
\affiliation{Marcus Center for Theoretical Chemistry, California Institute of Technology, Pasadena CA 91125, USA}

\author{Yongkang Wang}
\affiliation{Max Planck Institute for Polymer Research, Ackermannweg 10, 55128 Mainz, Germany}

\author{Christoph Schran}
\affiliation{Cavendish Laboratory, Department of Physics, University of Cambridge, Cambridge, CB3 0HE, UK}
\affiliation{Lennard-Jones Centre, University of Cambridge, Trinity Ln, Cambridge, CB2 1TN, UK}

\author{Mischa Bonn}
\affiliation{Max Planck Institute for Polymer Research, Ackermannweg 10, 55128 Mainz, Germany}

\author{Angelos Michaelides}
\email{am452@cam.ac.uk}
\affiliation{Yusuf Hamied Department of Chemistry, University of Cambridge, Lensfield Road, Cambridge, CB2 1EW, UK}
\affiliation{Lennard-Jones Centre, University of Cambridge, Trinity Ln, Cambridge, CB2 1TN, UK}

\author{Yair Litman}
\email{litmany@mpip-mainz.mpg.de}
\affiliation{Max Planck Institute for Polymer Research, Ackermannweg 10, 55128 Mainz, Germany}

\keywords{}

{\maketitle}

\tableofcontents

\onecolumngrid

\FloatBarrier

\newpage
\section{Additional simulation details} \label{sec:md_sims}
\subsection{Methods}

\textbf{Machine Learning Potential.} In this work, molecular dynamics simulations are performed using a committee of eight Behler–Parrinello neural network potentials (NNPs) developed in our previous work~\cite{kara_pairing_2024}.
The NNPs were trained on reference data from revPBE-D3(0) density functional theory~\cite{revpbed3_1, revpbed3_2} capturing both bulk and nanoconfined aqueous NaCl configurations across varying concentrations and confinement widths.
This has been shown to describe water–graphene interactions reliably~\cite{angelos_dft_water_2016, ondrej_revpbe_2017} as well as NaCl–water interaction~\cite{acs_nano_kara_2025}.

This training set explicitly includes interfacial environments similar to those studied here, such as the NaCl(aq)–graphene interface, allowing for the accurate modeling of ion–water and ion–surface interactions at the solid–liquid boundary.
To efficiently describe interatomic interactions, the models use atom-centered symmetry functions with a 12 Bohr cutoff, accounting for short-range interactions.
Long-range electrostatics are treated separately using a fixed-charge Coulomb baseline: +1 for cations, -1 for anions, 0 for carbon, and SPC/E charges for water.
The total system energy is partitioned as 
$E=E_{\textrm{sr}}+E_{\textrm{Coul}}$, where the NNP is trained only on the short-range component $E_{\textrm{sr}}$ and its associated forces.
During simulations, the full energy and forces are obtained by combining the NNP and Coulomb contributions.

We thoroughly validated the NNP against \textit{ab initio} molecular dynamics, showing excellent agreement in key structural and dynamical properties, including radial distribution functions, density profiles, and vibrational spectra~\cite{kara_pairing_2024}.
In addition, autocorrelation functions and computed bulk electrolyte conductivities align closely with experimental observations~\cite{acs_nano_kara_2025}, further confirming the model's reliability across relevant observables.

\textbf{Molecular Dynamics Simulations.} 
Molecular dynamics simulations were performed using the LAMMPS interface to the n2p2 NNP framework~\cite{n2p2_1, n2p2_2}.
All systems were constructed in orthorhombic simulation cells with periodic boundary conditions in all directions.
Each system consists of an aqueous NaCl solution confined between two parallel, rigid graphene sheets that form a slit pore.
The slit pore is wide enough to sustain bulk-like water density at its center (see Figure~\ref{fig:fig_dens_si}).
This setup enables the study of well-defined solid–liquid interfaces on both sides of the channel, facilitates statistical convergence, and allows direct comparison with previous slit-pore simulations~\cite{kara_pairing_2024}.
To eliminate spurious interactions along the $z$-axis (perpendicular to the interfaces), a vacuum layer at least three times the slit height was introduced, and the Yeh–Berkowitz correction \cite{yeh_berkowitz} was applied to account for the slab geometry properly.
Long-range electrostatics were treated using the particle–particle particle-mesh (PPPM) method.
Initial configurations were generated by randomly packing water molecules and NaCl ions between graphene sheets, followed by one ns of NVT equilibration with classical force fields, using the SPC/E water model, Dang ion parameters~\cite{dang_params}, and Werder carbon–water interactions~\cite{werder_params}.
This was followed by NVT equilibration of 25 ps using the NNP.
The equations of motion were integrated using a velocity-Verlet algorithm with a 0.5 fs time step, and temperature was maintained at 300 K via a Nosé–Hoover thermostat (damping time: 50 fs).
Production simulations were carried out in the NVE ensemble. For each state point, observables were averaged over 40 independent 200~ps trajectories to ensure robust statistics.
The reported error bars reflect the standard error across these replicas.
Across the four concentrations examined (pure water, 0.5, 1, and 2 M NaCl), this amounts to a cumulative sampling time exceeding 30 ns.
Forces were evaluated as the average over a committee of eight independently trained NNPs, ensuring robust statistical sampling.
We note that nuclear quantum effects were not included, as they are not expected to alter the observed trends~\cite{non_2024}.
However, they may influence the results semi-quantitatively, particularly with respect to the lifetime of dangling O--H bonds~\cite{wilkins_2015, wilkins_2017}

The NaCl-to-water ratios used in this work were approximately 1:110, 1:55, and 1:27.5, corresponding to bulk concentrations of 0.5 M, 1 M, and 2 M, respectively.
Each graphene sheet was constructed with lateral dimensions of 19.760\;\AA~by 21.390\;\AA.
To determine the appropriate slit height, and thereby the solution density, we performed additional equilibration runs in which one graphene layer was allowed to move along the $z$-axis under the influence of a piston exerting a pressure of 1 atm.
While the number of water molecules was kept fixed across all concentrations, the resulting slit heights varied slightly to accommodate the different salt contents.
The precise dimensions and number of molecules in each system are summarized in Table~\ref{tab:setup_details}.
We emphasize that although graphene sheets were mobile in the piston simulations, all carbon atoms were kept fixed during the production runs.

\textbf{Data Analysis.}
The VSFG spectra from molecular dynamics simulations were computed using the surface-specific velocity–velocity correlation function (ssVVCF) approach~\cite{ssvcf_2015}, applying a Hahn window of 1 ps.
The spectra represent an average over 40 independent 200-ps trajectories performed in the NVE ensemble, each initiated from uncorrelated configurations extracted from NVT simulations.
Throughout this work, we focus on the topmost interfacial water layer, which we show in Figure~\ref{fig:sfg_layers_si} to be the primary contributor to the VSFG response.
Including deeper layers, such as the second water layer, has minimal effect on the spectral features.
A central part of our analysis involves decomposing the interfacial response into contributions from three distinct water populations: H$_2$O-coordinated waters (not coordinating any ions), Na$^{+}$-coordinated waters, and Cl$^{-}$-coordinated waters.
A water molecule is considered coordinated to Na$^{+}$ or Cl$^{-}$ if its oxygen atom lies within 3.4\;\AA~or 4.0\;\AA~of the respective ion.
If a water molecule satisfies both criteria simultaneously, it is counted as coordinated to both ions.
These distance thresholds correspond to the first minima of the respective radial distribution functions.
We verified that moderate variations in these cutoffs do not qualitatively affect our conclusions.
Finally, to quantify the presence of dangling O--H bonds across different systems, we adopted the definition from Ref.~\citenum{yuki_free_oh_2018}.
In this framework, an interfacial O--H bond is considered as dangling if the distance between its oxygen atom and any other water oxygen (O$\cdots$O distance) exceeds 3.5\;\AA~ or the H--O$\cdots$O angle is greater than $50^\circ$.
Otherwise, the O--H bond is classified as hydrogen-bonded.
To compute the fraction of dangling O--H bonds, we follow previous studies~\cite{yuki_free_oh_2018, yuki_dft_free_oh_2019, chiang_2022, wang_ange_2024} and consider the sum of DA and DAA populations, where D and A represent the number of hydrogen-bond donors and acceptors per water molecule, respectively.

\newpage
\subsection{System Setup} 

\begin{figure*}[htp!]
    \centering
    \includegraphics[width=\textwidth]{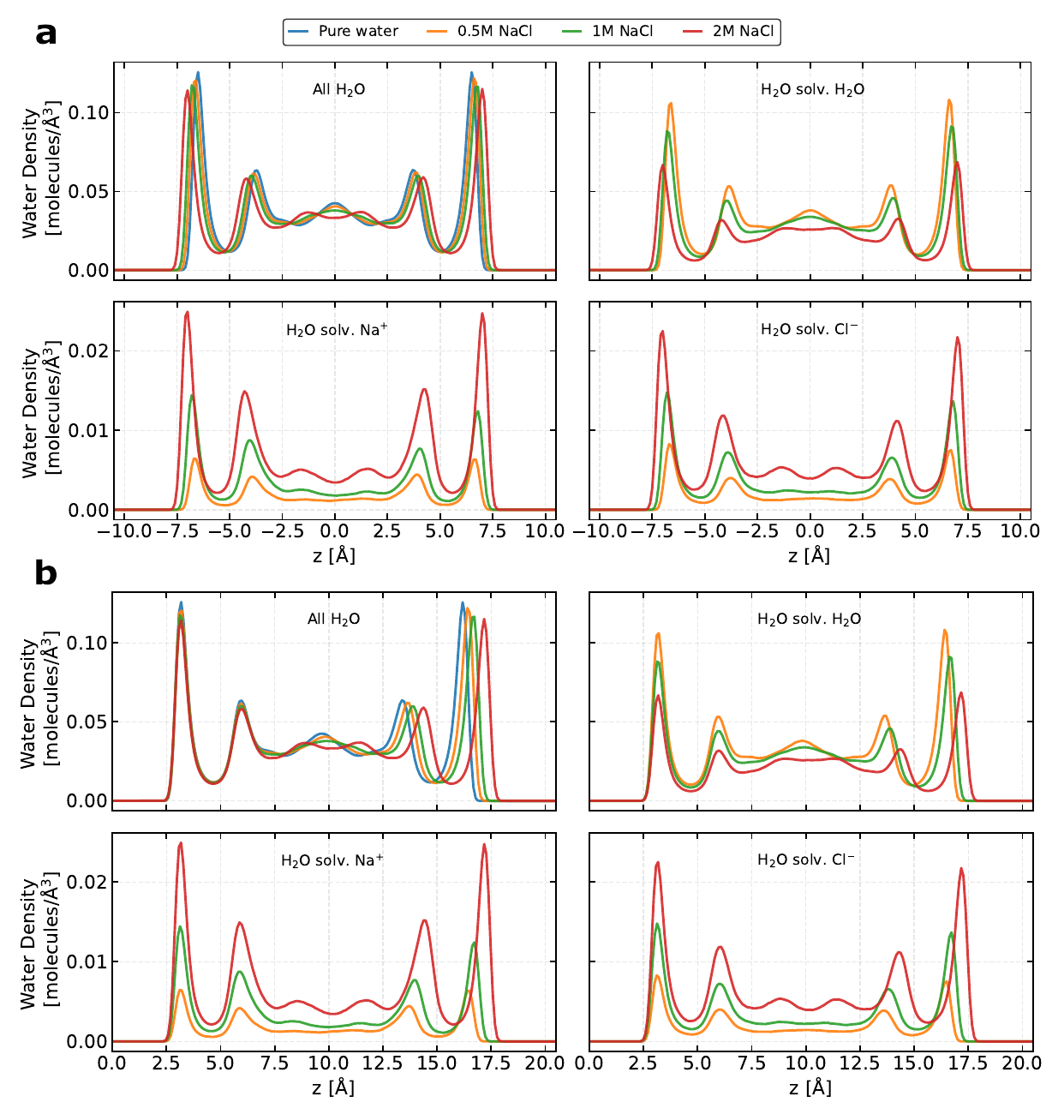}
    \caption{Density profiles of water across the slit pore, shown (a) symmetrized about $z=0$ and (b) referenced to a single graphene interface.
    Water molecules are classified into four categories: all waters, waters solvating other waters (not ions), waters solvating Na$^{+}$, and waters solvating Cl$^{-}$.}
    \label{fig:fig_dens_si}
\end{figure*}

We modeled aqueous NaCl solutions confined between two parallel rigid graphene sheets (19.760\;\AA~by 21.390\;\AA), forming slit pores wide enough to sustain bulk-like water density at the center (see Figure~\ref{fig:fig_dens_si}).
Salt-to-water ratios of approximately 1:110, 1:55, and 1:27.5 correspond to bulk concentrations of 0.5, 1, and 2 M, respectively.
The number of water molecules was kept constant across concentrations, with the slit height adjusted through piston equilibration runs, where one graphene sheet was allowed to move along the $z$-axis under 1 atm pressure.
This procedure yielded slightly different pore widths depending on the salt content.
Final system dimensions and compositions are reported in Table~\ref{tab:setup_details}.
During production simulations, all graphene atoms were held fixed.

\setlength{\tabcolsep}{10pt}
\begin{table}[htp!]
\centering
\caption{Overview of the systems considered in this work.
For each system, we report their slit width, $\textrm{W}$; the graphene dimensions; the number of water molecules, $N_{\textrm{H}_2\textrm{O}}$; the number of Na$^{+}$ ions, $N_{\textrm{Na$^{+}$}}$; the number of Cl$^{-}$ ions, $N_{\textrm{Cl$^{-}$}}$; the concentration; the number of independent runs, $N_{\textrm{runs}}$; and the simulation production time, $t_{\textrm{sim}}$.}
\vspace{0.4cm}
\label{tab:setup_details}
\begin{tabular}{@{}cccccccc@{}}
\toprule
$\textrm{W}$ & \begin{tabular}[c]{@{}c@{}}Graphene\\ dims. [\AA $\times$\AA]\end{tabular} & $N_{\textrm{H}_2\textrm{O}}$ & $N_{\textrm{Na$^{+}$}}$ & $N_{\textrm{Cl$^{-}$}}$ &  Conc. [M] & $N_{\textrm{runs}}$ & $t_{\textrm{sim}}$ [ps]\\ \midrule
19.36 & 19.760 $\times$ 21.390 & 232 & 0 & 0 & 0 & 40 & 200 \\
19.61 & 19.760 $\times$ 21.390 & 232 & 2 & 2 & 0.5 & 40 & 200 \\
19.85 & 19.760 $\times$ 21.390 & 232 & 4 & 4 & 1 & 40 & 200 \\
20.35 & 19.760 $\times$ 21.390 & 232 & 8 & 9 & 2 & 40 & 200 \\ \bottomrule
\end{tabular}%
\end{table}

\newpage
The corresponding water density profiles across the slit pores are shown in Figure~\ref{fig:fig_dens_si}. 
As illustrated in Figure~\ref{fig:fig_dens_si}b, the interfacial water molecules were defined as those located between the graphene surface and the first minimum of the water oxygen density profile, found at approximately 4.5~\AA{} from the surface.

\newpage

\begin{figure*}[htp!]
    \centering
    \includegraphics[width=\textwidth]{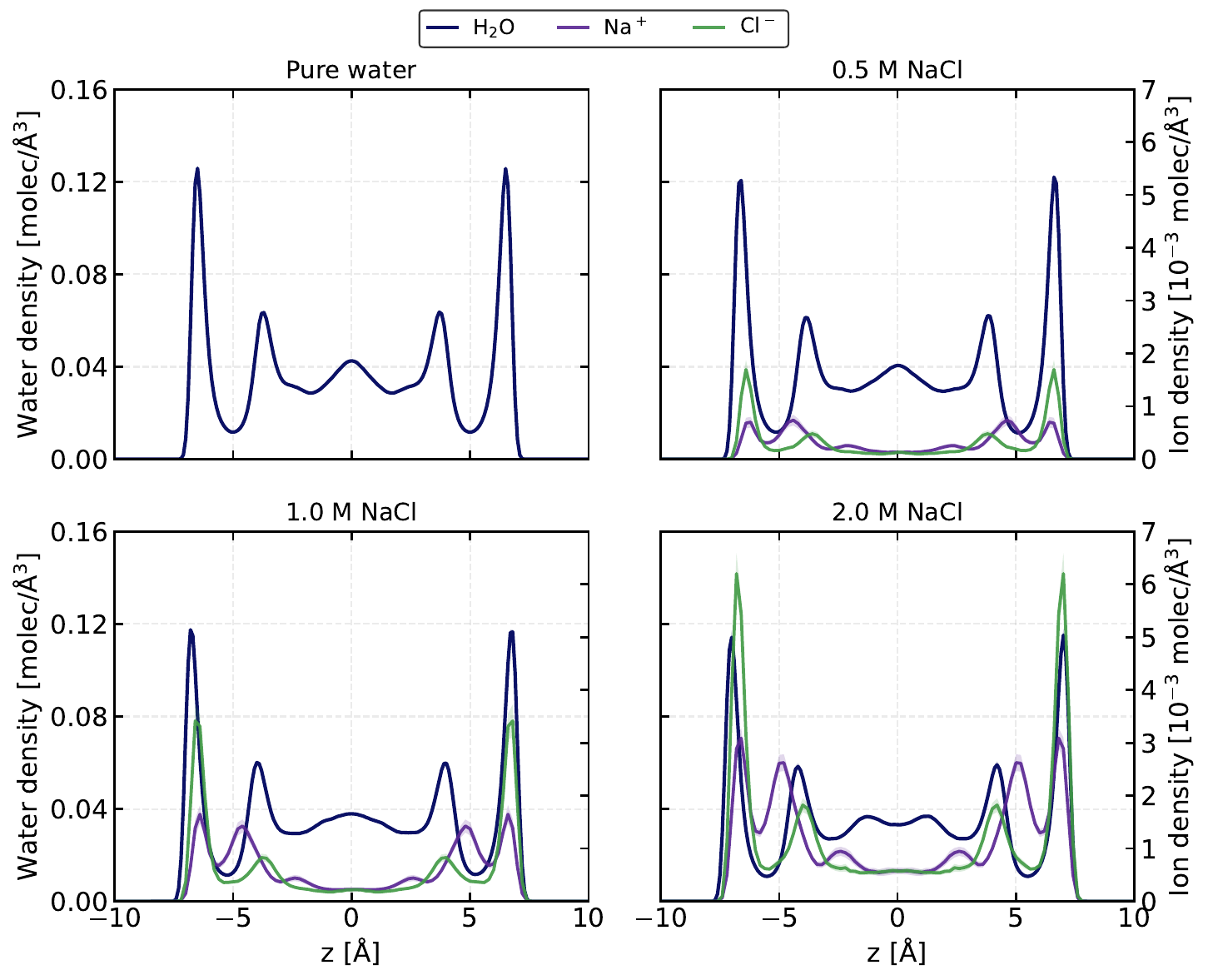}
    \caption{Symmetrized density profiles of water and ions across the slit pore for the different NaCl concentrations studied. The total water density is shown together with the corresponding Na$^+$ and Cl$^-$ ion densities, all symmetrized about $z = 0$.}
    \label{fig:fig_dens_si_ions}
\end{figure*}

For completeness, in Figure~\ref{fig:fig_dens_si_ions} we show the total water density profile accompanied by the ion density.

\begin{figure*}[htp!]
    \centering
    \includegraphics[width=\textwidth]{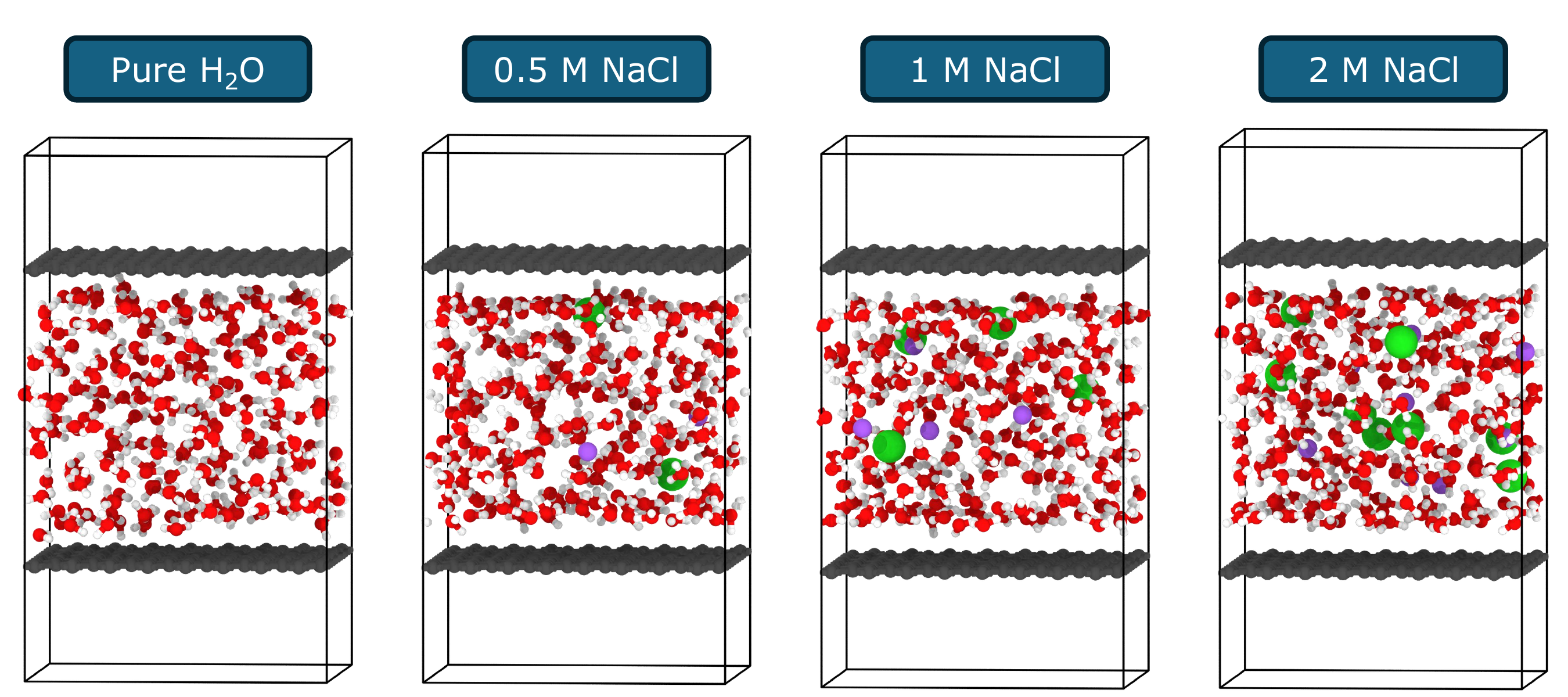}
    \caption{Representative simulation snapshots of the slit-pore systems studied.}
    \label{fig:fig_systems_si}
\end{figure*}

\newpage
Representative snapshots of the NaCl solutions considered in this work are shown in Figure~\ref{fig:fig_systems_si}.

\newpage

\section{Additional system characterization}

\begin{figure*}[htp!]
    \centering
    \includegraphics[width=\textwidth]{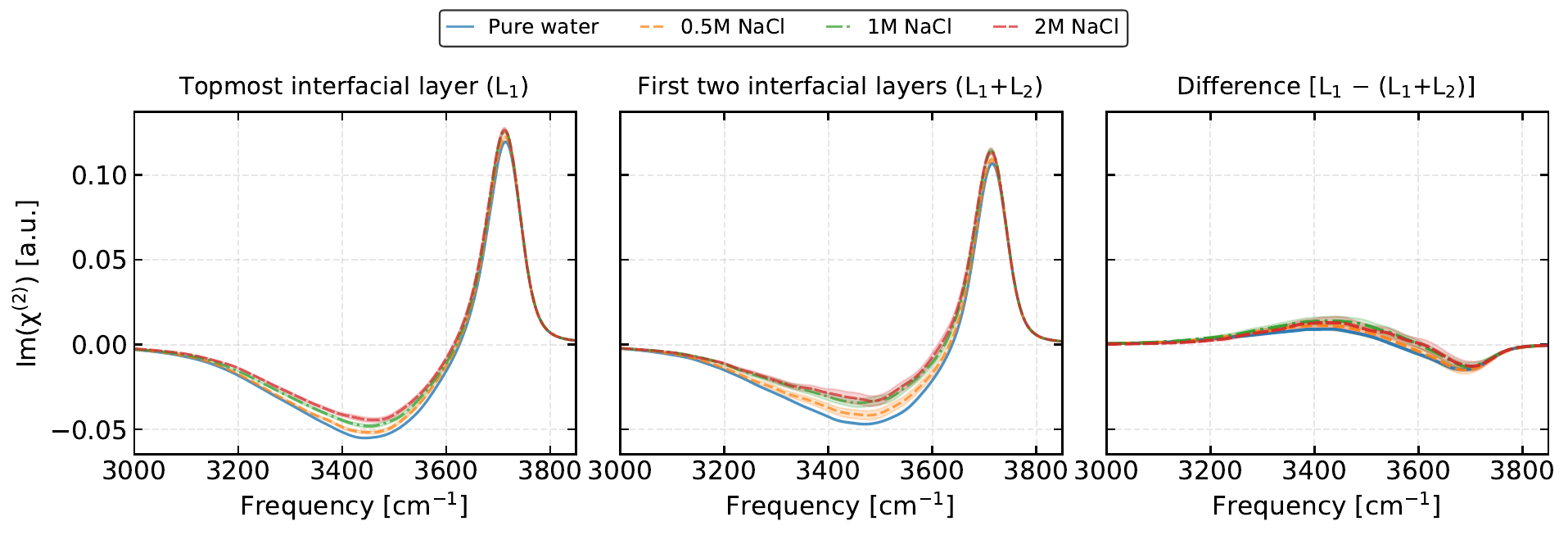}
    \caption{VSFG spectra of interfacial water at the graphene–water interface.
    (Left) Contribution from the topmost interfacial layer, L$_1$.
    (Center) Contribution from the first two interfacial layers, L$_{1}+$L$_{2}$).
    (Right) Difference spectrum, highlighting the marginal influence of the second layer on the overall response.
    Results are shown for pure water and NaCl solutions at concentrations of 0.5~M, 1~M, and 2~M.}
    \label{fig:sfg_layers_si}
\end{figure*}

In the main text, we focus on the topmost interfacial water layer, which provides the dominant contribution to the spectra, while the second layer influences the response only marginally.
Figure~\ref{fig:sfg_layers_si} illustrates this by comparing VSFG spectra computed from only the topmost layer with those including the first two layers.

\begin{figure*}[htp!]
    \centering
    \includegraphics[width=\textwidth]{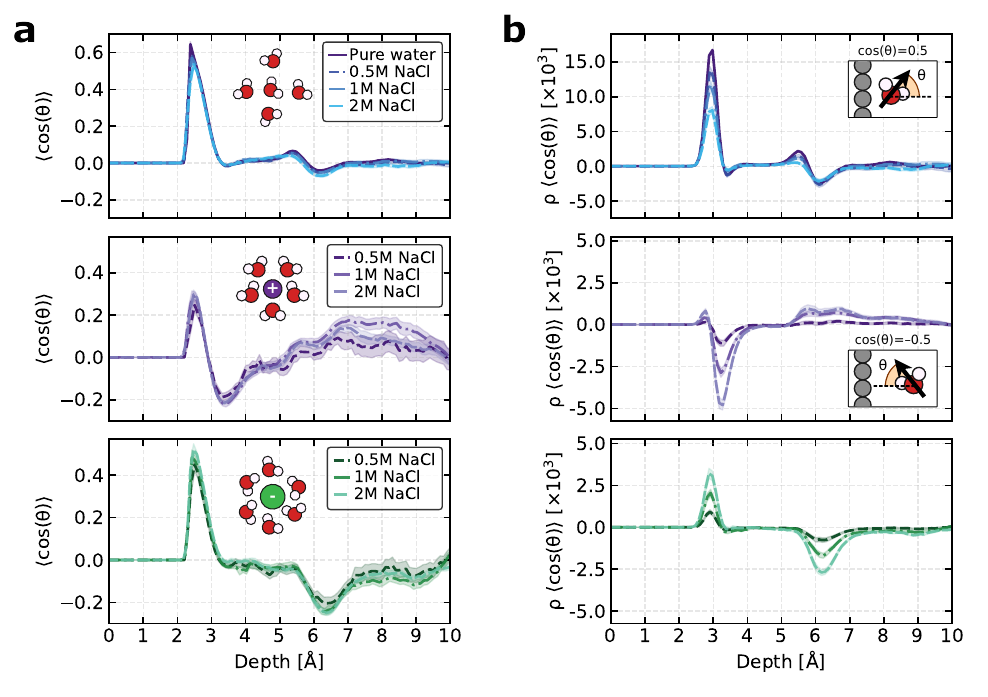}
    \caption{(a) Average orientation of the H–O–H bisector of water molecules relative to the surface normal, and (b) product of the water density profiles and the average orientation.
    Results are shown separately for free water molecules (blue shades), water molecules coordinating Na$^{+}$ ions (purple shades), and those coordinating Cl$^{-}$ ions (green shades).
    A positive (negative) $\langle \cos(\theta)\rangle$ value indicates net orientation towards the bulk (towards the graphene surface).
    }
    \label{fig:rhocos_si}
\end{figure*}

\newpage
In Figure~\ref{fig:rhocos_si}, we present the orientational profiles of the different water classes, along with the same profiles weighted by their density.
These weighted profiles act as a proxy for understanding VSFG variations and show agreement with the observations reported in the main text.

\newpage

\begin{figure*}[htp!]
    \centering
    \includegraphics[width=\textwidth]{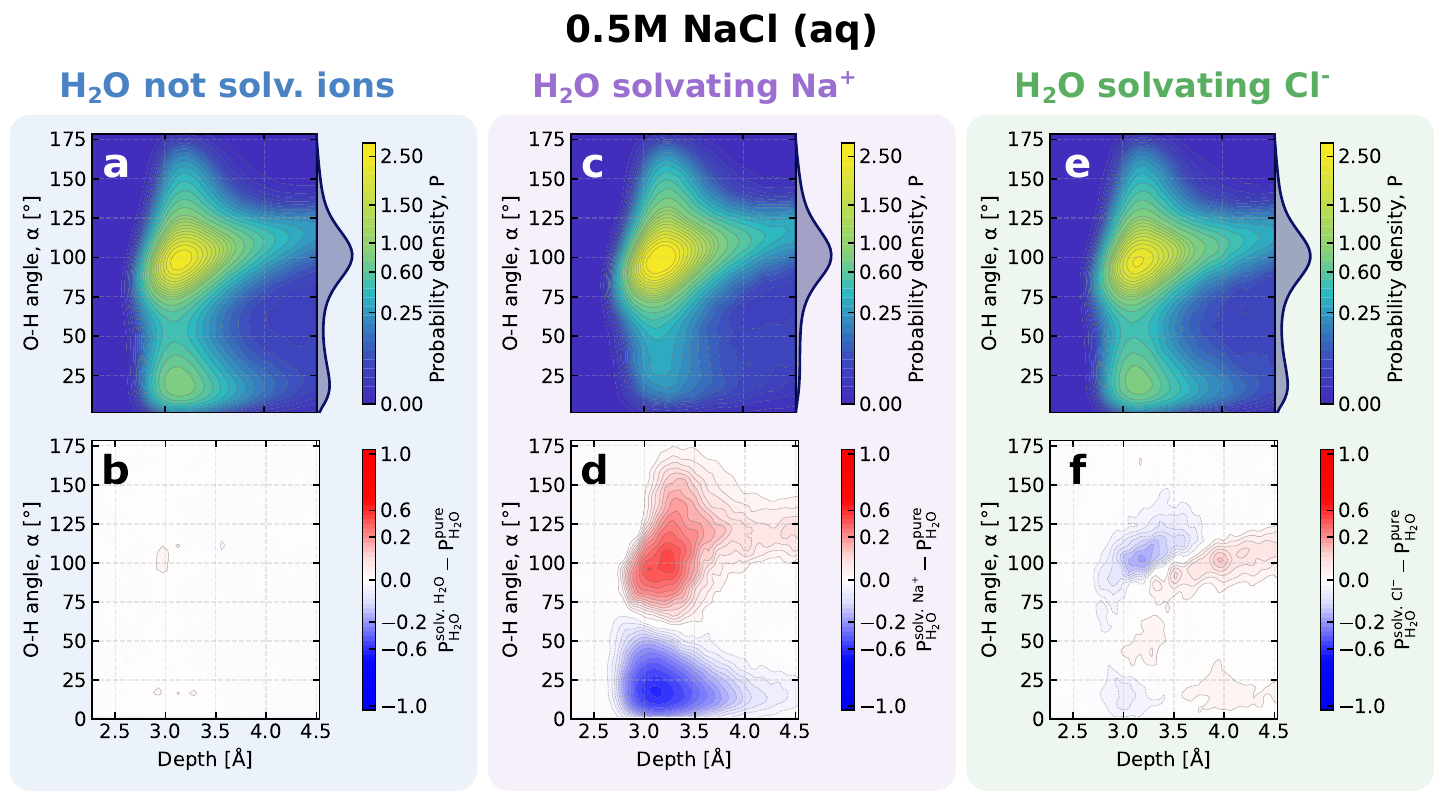}
    \caption{Microscopic analyses of the graphene–NaCl(aq) interface for a 0.5~M NaCl solution.
    (a) Probability distribution of O–H bond orientations in water molecules not solvating ions as a function of their depth from the graphene interface and their angle relative to the surface normal.
    (b) Difference in the probability distribution shown in (a) relative to that of pure water at the graphene interface.
    Positive values indicate features that appear compared to pure water, while negative values indicate features that disappear compared to pure water.
    (c–d) Same as (a–b) but for water molecules solvating Na$^{+}$.
    (e–f) Same as (a–b) but for water molecules solvating Cl$^{-}$.
    }
    \label{fig:fig_micros_05}
\end{figure*}

\begin{figure*}[htp!]
    \centering
    \includegraphics[width=\textwidth]{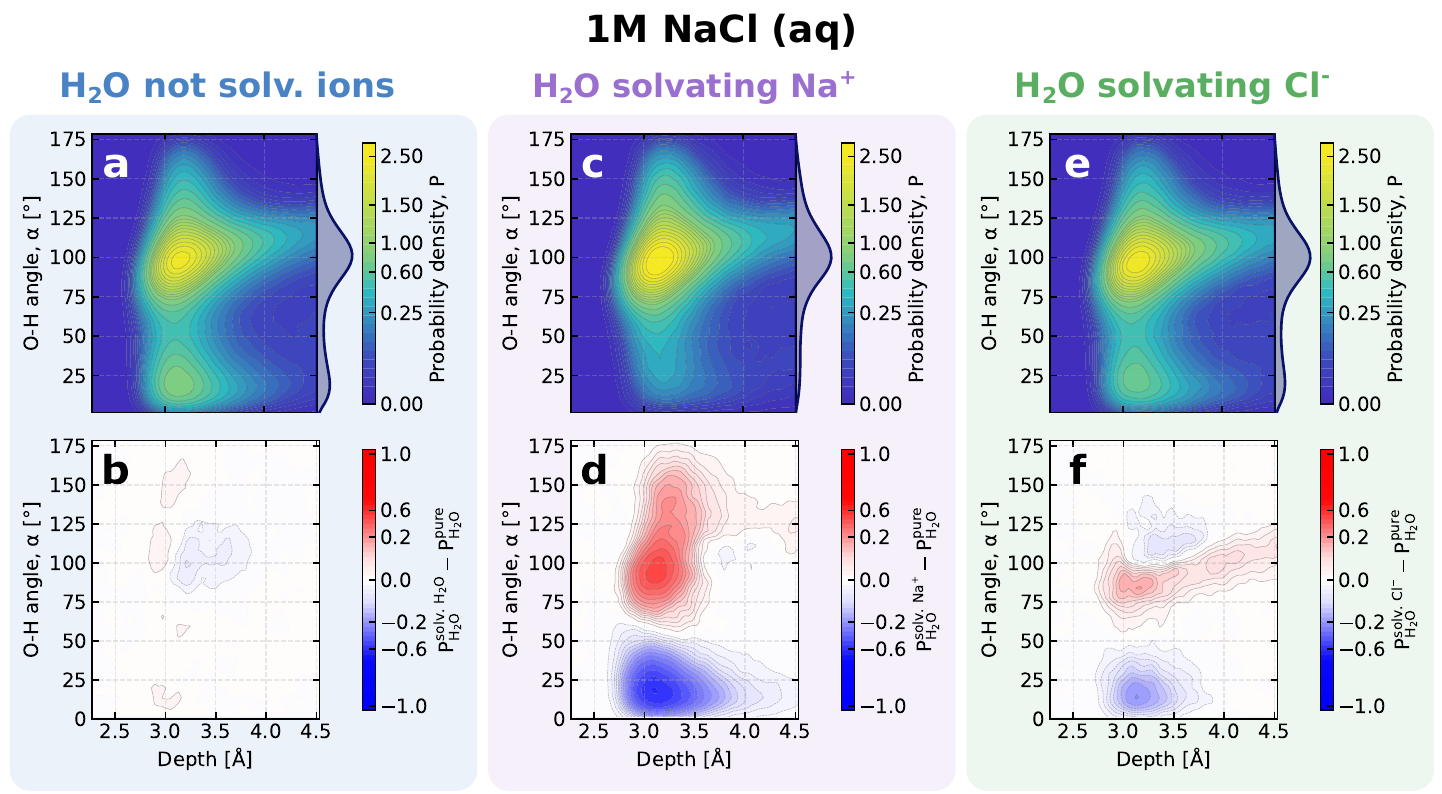}
    \caption{Microscopic analyses of the graphene–NaCl(aq) interface for a 1~M NaCl solution.
    (a) Probability distribution of O–H bond orientations in water molecules not solvating ions as a function of their depth from the graphene interface and their angle relative to the surface normal.
    (b) Difference in the probability distribution shown in (a) relative to that of pure water at the graphene interface.
    Positive values indicate features that appear compared to pure water, while negative values indicate features that disappear compared to pure water.
    (c–d) Same as (a–b) but for water molecules solvating Na$^{+}$.
    (e–f) Same as (a–b) but for water molecules solvating Cl$^{-}$.
    }
    \label{fig:fig_micros_1}
\end{figure*}

In Figures~\ref{fig:fig_micros_05} and~\ref{fig:fig_micros_1}, we present microscopic analyses of the graphene–NaCl(aq) interface for 0.5 and 1~M solutions, complementing the 2~M case discussed in the main text.

\newpage

\newpage

\section{Dangling O--H characterization}

\begin{figure*}[htp!]
    \centering
    \includegraphics[width=\textwidth]{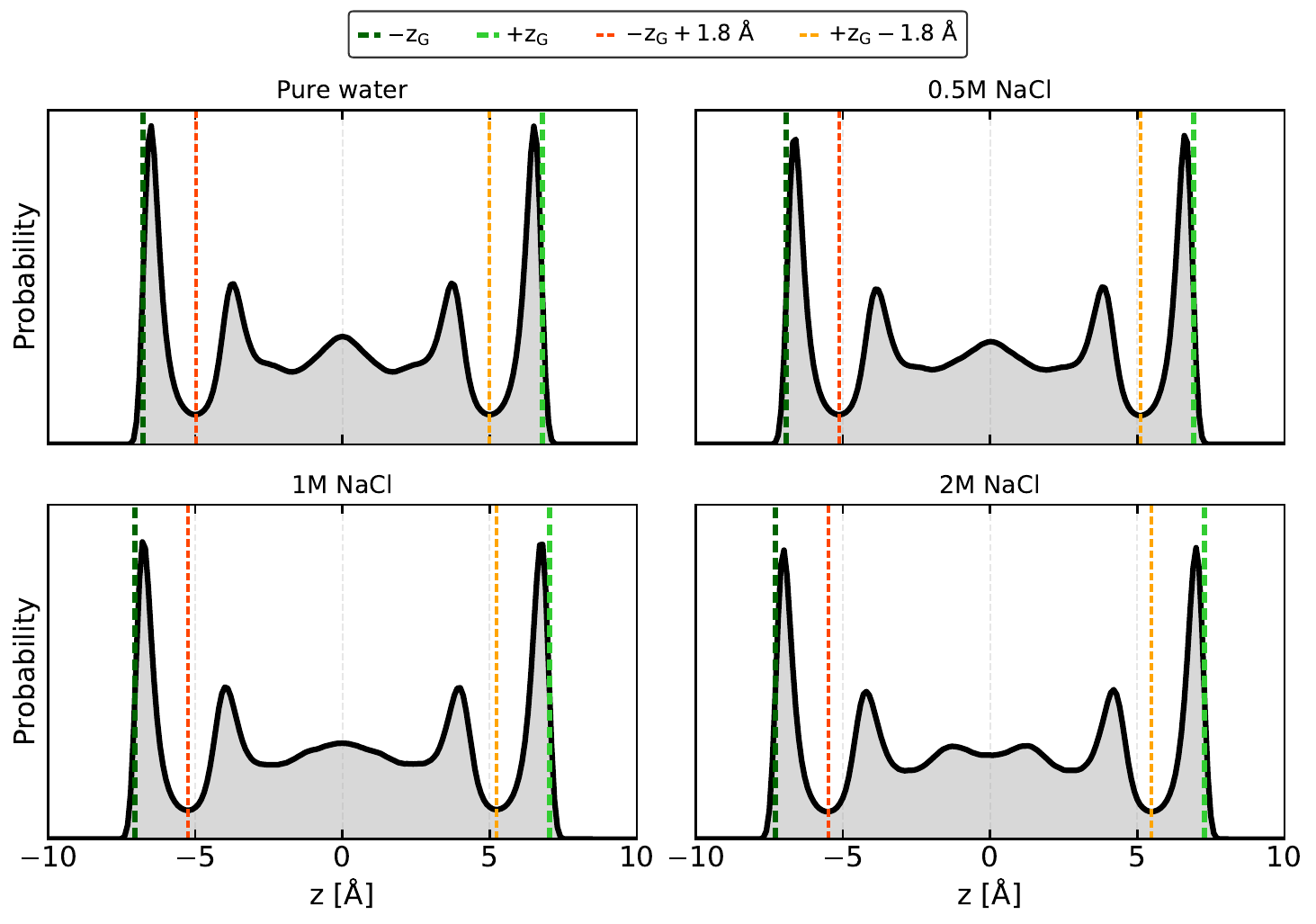}
    \caption{Density profiles across the slit pore along the surface normal ($z$) for pure water and NaCl solutions at different concentrations.
    The Gibbs dividing surfaces ($\pm z_G$) and the boundaries used to define the interfacial region ($-z_G+1.8$ \AA\; and $+z_G-1.8$ \AA, orange lines).}
    \label{fig:fig_dens_gds}
\end{figure*}

An interfacial O–H bond is classified as `free' if the distance between its oxygen atom and any other water oxygen (O$\cdots$O) exceeds 3.5\;\AA\; and the D–O$\cdots$O angle is greater than $50^\circ$; otherwise, it is considered hydrogen-bonded.
This definition follows Ref.~\citenum{yuki_free_oh_2018}.
The fraction of H$_2$O molecules carrying free O–H bonds is obtained by summing the DA and DAA populations, where D and A denote the number of hydrogen-bond donors and acceptors per water molecule, respectively.
This analysis is restricted to the interfacial region, defined as
$$
z \leq -z_G + 1.8~\text{\AA} \quad \text{or} \quad z \geq +z_G - 1.8~\text{\AA},
$$
with $z_G$ the $z$-coordinate of the Gibbs dividing surface.
The definition of the interfacial region and the position of $z_G$ are illustrated in Figure~\ref{fig:fig_dens_gds}.

\subsection*{Ion Coordination Enhances Dangling O--H Orientation at the Interface}

\begin{figure*}[htp!]
    \centering
    \includegraphics[width=0.45\textwidth]{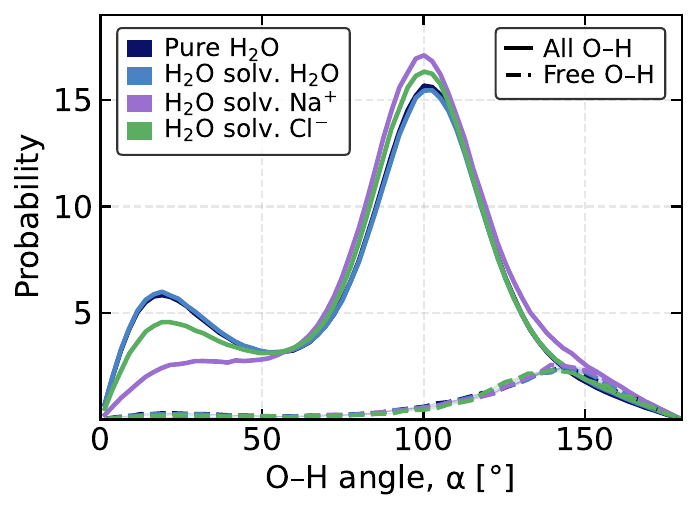}
    \caption{Orientational distributions of all interfacial O--H bonds (solid lines) and free O--H bonds only (dashed lines) for pure water, H$_{2}$O-solvating waters, Na$^{+}$-solvating waters, and Cl$^{-}$-solvating waters in a 2~M NaCl solution.}
    \label{fig:fig_oh_distro}
\end{figure*}

Here we examine how salt influences the population of dangling O--H bonds at the graphene–water interface.
To identify such species, as previously introduced, we adopt the well-established definition of dangling O--H bonds based on the absence of hydrogen bonding interactions, following the criteria of Ref.~\citenum{yuki_free_oh_2018}.
While the 2D orientation–depth maps in Figure 4 provide rich structural insight, they are less informative for detecting dangling O--H bonds due to their relatively low abundance, as evidenced in Figure~\ref{fig:fig_oh_distro}.

\begin{figure}
    \centering
    \includegraphics[width=0.48\textwidth]{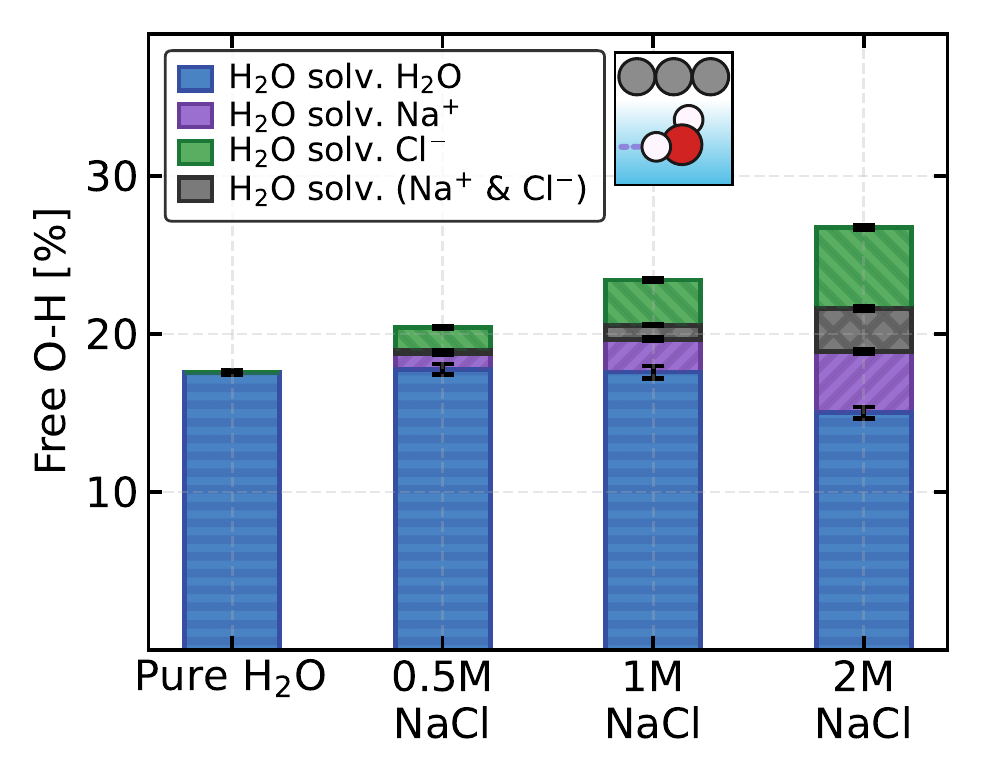}
    \caption{\textbf{Dangling O--H bond characterization at the graphene–NaCl(aq) interface.}
    Fraction of dangling O--H bonds among interfacial (first-layer) water molecules as a function of NaCl concentration, separated into contributions from H$_2$O-solvating waters, Na$^{+}$-solvating waters (not coordinating Cl$^{-}$), Cl$^{-}$-solvating waters (not coordinating Na$^{+}$), and waters simultaneously coordinating both ions.
    The schematic illustrates a dangling O--H bond of an interfacial water molecule and a hydrogen-bonded O--H bond, which give rise to distinct signatures in the VSFG spectra.
    }
    \label{fig:fig_oh}
\end{figure}

To quantify how salt and its concentration affect interfacial structure, we calculate the fraction of interfacial O--H bonds considered to be dangling bonds within each water environment.
Figure~\ref{fig:fig_oh} shows this dangling O--H fraction as a function of concentration, resolved by water type.

\newpage

\begin{figure*}[htp!]
    \centering
    \includegraphics[width=0.6\textwidth]{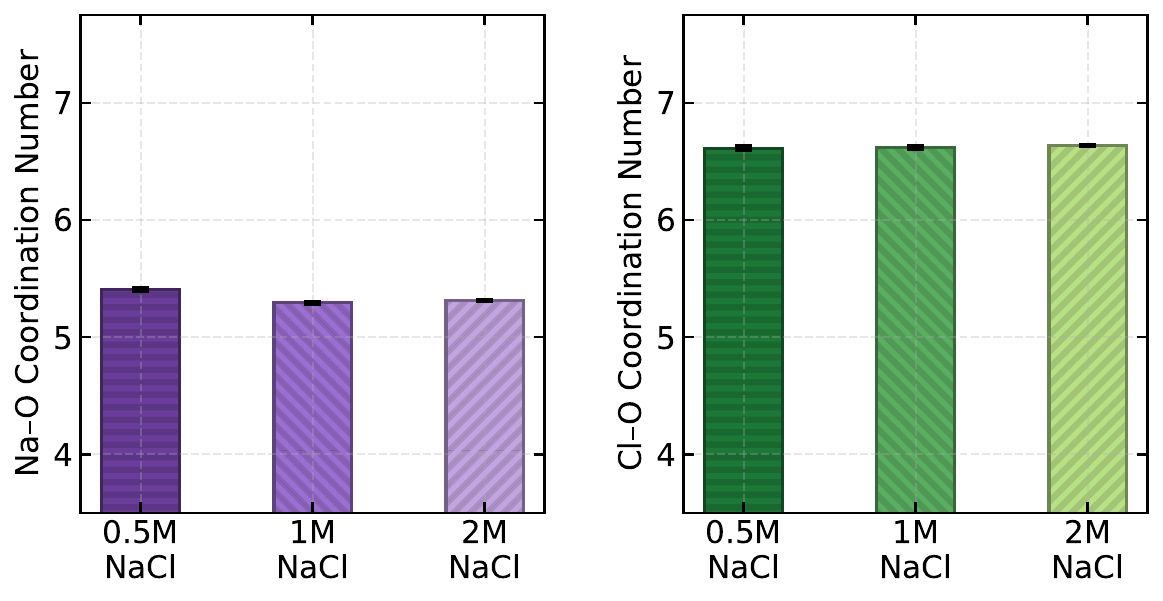}
    \caption{
    Average Na–O and Cl–O coordination numbers at different NaCl concentrations (0.5~M, 1~M, and 2~M).}
    \label{fig:coord-si}
\end{figure*}

As it can be seen in Figure~\ref{fig:fig_oh}, across all concentrations we observe an overall increase in the proportion of dangling O--H bonds upon salt addition.
When normalized with respect to the total number of interfacial water molecules, water solvating Cl$^{-}$ appears to contribute a comparable or even larger fraction than those solvating Na$^{+}$.
This reflects the larger hydration shell of Cl$^{-}$~\cite{acs_nano_kara_2025}, meaning that each anion coordinates more interfacial waters, rather than a higher propensity of individual Cl$^{-}$-bound waters to expose dangling O--H bonds (see Figure~\ref{fig:coord-si}).

\newpage

\begin{figure*}[htp!]
    \centering
    \includegraphics[width=0.6\textwidth]{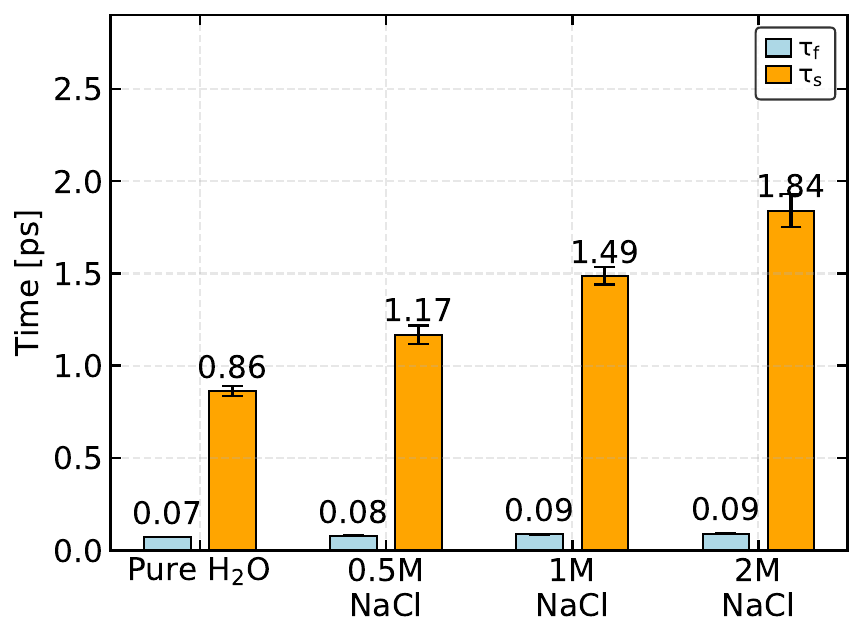}
    \caption{Lifetimes of free O–H bonds at the graphene–water interface as a function of NaCl concentration.
    Bars show the fitted fast ($\tau_{f}$) and slow ($\tau_{s}$) components obtained from biexponential fits of the correlation function $C(t)$.
    Error bars denote the fitting uncertainties.}
    \label{fig:free_oh_lifetime}
\end{figure*}

When the data are instead normalized by water type (Figure 4), Na$^{+}$-coordinated waters indeed exhibit a higher dangling O--H fraction.
This trend parallels the enhancement of the dangling O--H peak in the VSFG spectra (Figure 3b,c), confirming that the spectral changes reflect genuine structural differences at the interface.
Notably, in 1~M and 2~M solutions, the dangling O--H population at the graphene–electrolyte interface is higher than that reported for the air–water interface (about 24\% \cite{yuki_dft_free_oh_2019}).
To probe this effect further, we compute the lifetimes of dangling O--H bonds, which are experimentally accessible via pump–probe VSFG spectroscopy~\cite{bonn_2013_free_oh}.
For this, we calculated the lifetime of free O--H bonds using the time correlation function
\[
C(t) = \frac{\langle n(0)\,n(t) \rangle}{\langle n(0) \rangle},
\]
where $n(t) = 1$ if an O--H bond is classified as free at time $t$, and $n(t) = 0$ otherwise.
The term $\langle n \rangle$ denotes the ensemble average of $n$.
We can now fit $C(t)$ with a double exponential of the form\cite{lifetime_oh},
$$
C(t) = a\, e^{-\tfrac{t}{\tau_f}} + b\, e^{-\tfrac{t}{\tau_s}} + c,
$$
where $a$, $b$, and $c$ are fitting coefficients.
The time constants $\tau_f$ and $\tau_s$ represent the fast and slow components, respectively.
$\tau_f$ reflects the librational motion of water, while $\tau_s$ corresponds to the reorientation of the free O--H bond as it rotates and eventually forms a hydrogen bond with another water molecule at the interface.
As shown in Figure~\ref{fig:free_oh_lifetime}, these lifetimes increase with salt concentration.

\newpage

At the air–water interface, the lifetime is about 1.06~ps~\cite{yuki_dft_free_oh_2019}, whereas at the graphene–water interface studied here, it ranges from 0.86~ps in pure water to 1.84~ps at 2~M NaCl, highlighting the distinct dynamical response of dangling O--H bonds at solid–water interfaces.
As spectral shapes alone can be ambiguous, lifetime measurements provide a readily accessible experimental signature and more robust route to disentangling interfacial structural and dynamical effects, thereby establishing a benchmark for future experimental validation.

\newpage

\section{Hydrogen-bond network topology and connectivity}

\begin{figure*}[htp!]
    \centering
    \includegraphics[width=0.6\textwidth]{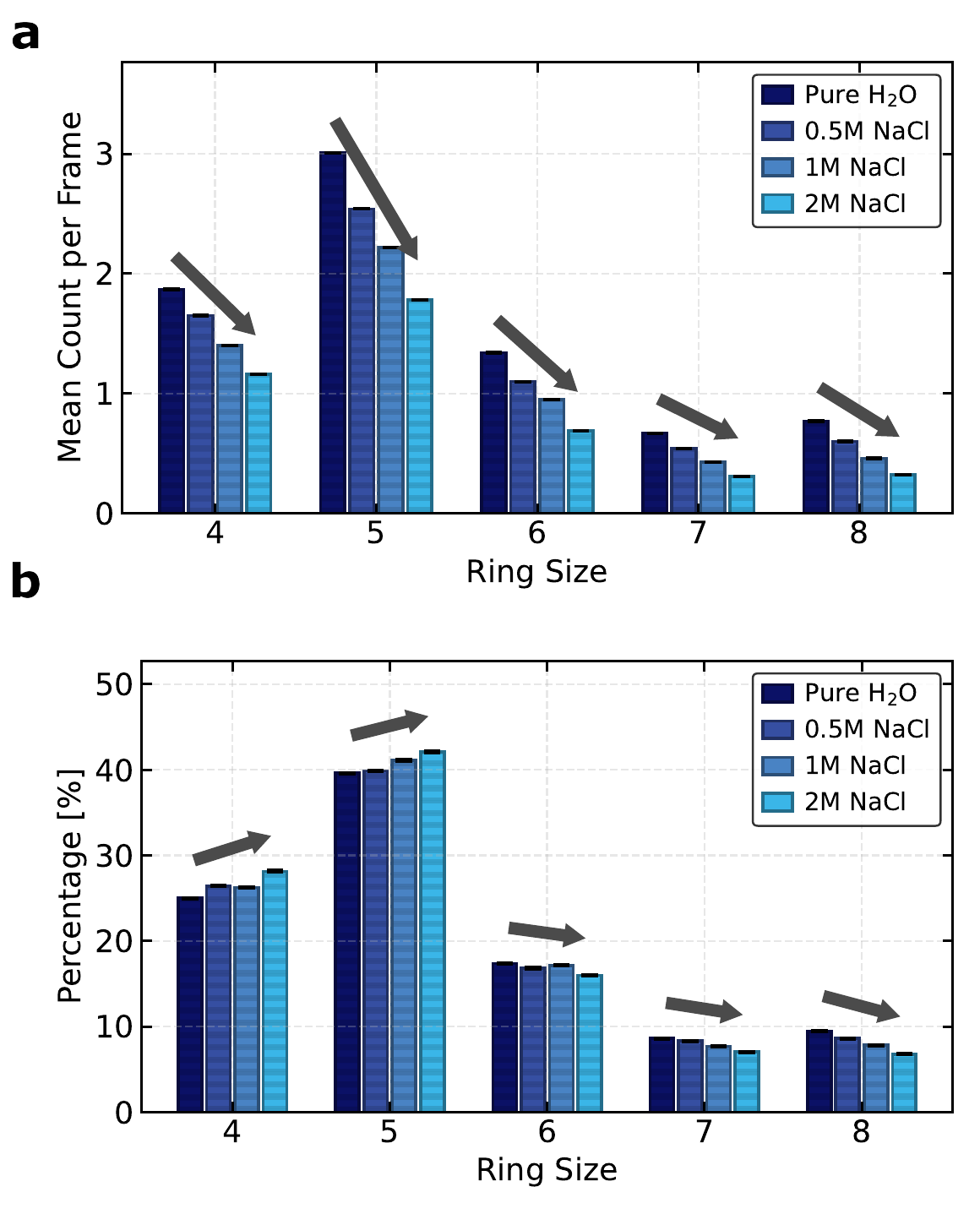}
    \caption{(a) Mean count per frame and (b) distribution of hydrogen-bonded ring sizes in interfacial water for pure water and NaCl solutions at 0.5~M, 1~M, and 2~M.}
    \label{fig:ring_percentage_si}
\end{figure*}

In the main text, we described how salt disrupts the extended hydrogen-bond network at the graphene–water interface, breaking large two-dimensional structures.
Figure~\ref{fig:ring_percentage_si}a confirms that, upon the addition of salt, fewer rings are formed.
Figure~\ref{fig:ring_percentage_si}b further shows a subtle shift in the distribution toward smaller chains and rings.

\newpage

\begin{figure*}[htp!]
    \centering
    \includegraphics[width=0.4\textwidth]{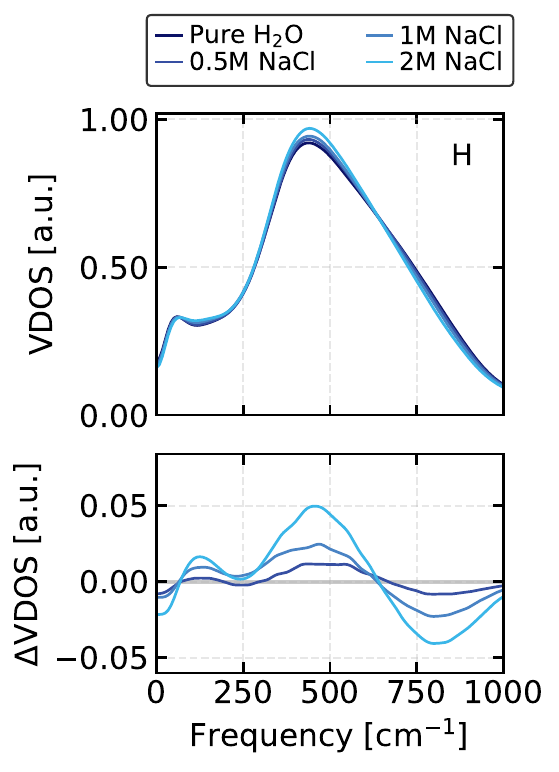}
    \caption{Low-frequency interfacial vibrational density of states (VDOS) for H as a function of NaCl concentration, with $\Delta$VDOS shown relative to pure H$_2$O.}
    \label{fig:vdos_h}
\end{figure*}

\newpage
\bibliography{references.bib}